\documentclass{aa}  

\usepackage{float}
\usepackage[caption = false]{subfig}
\usepackage{graphicx}
\usepackage{xcolor}
\usepackage[varg]{txfonts}
\usepackage[breaklinks=true,colorlinks=true,urlcolor=blue,linkcolor=blue,citecolor=blue,bookmarks=true,bookmarksopen=true]{hyperref}
\usepackage{siunitx}
\usepackage{amsmath, mathrsfs}
\usepackage{mathtools}
\usepackage{upgreek}
\usepackage{makecell} 
\DeclareMathAlphabet{\mathbbmsl}{U}{bbm}{m}{sl}

\newcolumntype{L}[1]{>{\raggedright\arraybackslash}p{#1}}
\newcolumntype{C}[1]{>{\centering\arraybackslash}p{#1}}
\newcolumntype{R}[1]{>{\raggedleft\arraybackslash}p{#1}}

\usepackage{natbib,twoopt}
\bibpunct{(}{)}{;}{a}{}{,} 					
\makeatletter
\newcommandtwoopt{\citeads}[3][][]{\href{http://adsabs.harvard.edu/abs/#3}%
{\def\hyper@linkstart##1##2{}%
\let\hyper@linkend\@empty\citetlp[#1][#2]{#3}}}
\newcommandtwoopt{\citepads}[3][][]{\href{http://adsabs.harvard.edu/abs/#3}%
{\def\hyper@linkstart##1##2{}%
\let\hyper@linkend\@empty\citep[#1][#2]{#3}}}
\newcommandtwoopt{\citetads}[3][][]{\href{http://adsabs.harvard.edu/abs/#3}%
{\def\hyper@linkstart##1##2{}%
\let\hyper@linkend\@empty\citet[#1][#2]{#3}}}
\newcommandtwoopt{\citeyearads}[3][][]%
{\href{http://adsabs.harvard.edu/abs/#3}
{\def\hyper@linkstart##1##2{}%
\let\hyper@linkend\@empty\citeyear[#1][#2]{#3}}}
\makeatother

\begin{document} 

\title{Polarization-dependent beam shifts upon metallic reflection in high-contrast imagers and telescopes}

\subtitle{}


\authorrunning{R.\,G. van Holstein et al.}

\author{R.\,G.~van~Holstein\inst{\ref{inst:esosantiago}} 
\and C.\,U.~Keller\inst{\ref{inst:lowell},\ref{inst:leiden}} 
\and F.~Snik\inst{\ref{inst:leiden}} 
\and S.\,P.~Bos\inst{\ref{inst:leiden}} 
}

\institute{
European Southern Observatory, Alonso de C\'{o}rdova 3107, Casilla 19001, Vitacura, Santiago, Chile \newline
e-mail: \texttt{rob.vanholstein@eso.org} \label{inst:esosantiago}
\and Lowell Observatory, 1400 W Mars Hill Road, Flagstaff, AZ, USA \label{inst:lowell}
\and Leiden Observatory, Leiden University, PO Box 9513, 2300 RA Leiden, The Netherlands \label{inst:leiden}
}

\date{Received 19 January 2022 / Accepted 11 July 2023}

\abstract{
To directly image rocky exoplanets in reflected (polarized) light, future space- and ground-based high-contrast imagers and telescopes aim to reach extreme contrasts at close separations from the star.
However, the achievable contrast will be limited by reflection-induced polarization aberrations.
While polarization aberrations can be modeled with numerical codes, these computations provide little insight into the full range of effects, their origin and characteristics, and possible ways to mitigate them.
}
{
We aim to understand polarization aberrations produced by reflection off flat metallic mirrors at the fundamental level.
}
{
We used polarization ray tracing to numerically compute polarization aberrations and interpret the results in terms of the polarization-dependent spatial and angular Goos-H\"anchen and Imbert-Federov shifts of the beam of light as described with closed-form mathematical expressions in the physics literature.
}
{
We find that all four beam shifts are fully reproduced by polarization ray tracing. 
We study the origin and characteristics of the shifts as well as the dependence of their size and direction on the beam intensity profile, incident polarization state, angle of incidence, mirror material, and wavelength.
Of the four beam shifts, only the spatial Goos-H\"anchen and Imbert-Federov shifts are relevant for high-contrast imagers and telescopes because these shifts are visible in the focal plane and create a polarization structure in the point-spread function that reduces the performance of coronagraphs and the polarimetric speckle suppression close to the star.
}
{
Our study provides a fundamental understanding of the polarization aberrations resulting from reflection off flat metallic mirrors in terms of beam shifts and lays out the analytical and numerical tools to describe these shifts.
The beam shifts in an optical system can be mitigated by keeping the f-numbers large and angles of incidence small.
Most importantly, mirror coatings should not be optimized for maximum reflectivity, but should be designed to have a retardance close to $180\degr$.
The insights from our study can be applied to improve the performance of SPHERE-ZIMPOL at the VLT and future telescopes and instruments such as the Roman Space Telescope, the Habitable Worlds Observatory, GMagAO-X at the GMT, PSI at the TMT, and PCS (or EPICS) at the ELT.
}


\keywords{Polarization --
		   Telescopes --
		   Instrumentation: high angular resolution --
		   Instrumentation: polarimeters --
		   Methods: analytical --
		   Methods: numerical
}

\maketitle

%
%

\section{Introduction}
\label{sec:introduction}

To directly image rocky exoplanets in (polarized) reflected visible and near-infrared light, future space telescopes and extremely large ground-based telescopes and instruments aim to reach extreme planet-to-star contrast ratios at diffraction-limited angular separations from the star. 
Even though the optical systems of these high-contrast imagers will minimize scalar aberrations, the coronagraphic performance and achievable contrast will still be limited by polarization aberrations \citep[e.g.,][]{chipman1989polarization, mcguire1990diffraction, sanchez1992instrumental, mcguire1994polarization1, mcguire1994polarization2, breckinridge2015polarization}.
Polarization aberrations are minute, polarization-dependent variations of the amplitude and phase of the electromagnetic field across a beam of light that result in a polarization structure in the point-spread function (PSF). 
Polarization aberrations are predominantly caused by reflection off oblique and/or curved metallic mirrors and originate directly from the Fresnel reflection coefficients.
The first-order polarization aberrations, that is, the sub-wavelength, polarization-dependent shifts of the beam of light, most negatively affect the achievable contrast.
Because polarization aberrations are different for orthogonal polarization components of unpolarized light, adaptive optics cannot fully correct these aberrations \citep{breckinridge2015polarization}.

Recently, it has become clear that high-angular-resolution polarimeters are also affected by polarization aberrations.
The polarization aberrations of the Gemini South telescope appear to be limiting the polarimetric contrast achieved by the Gemini Planet Imager at the smallest angular separations from the star \citep{millar2022polarization}.
Moreover, the polarimetric speckle suppression of the high-contrast imaging polarimeter SPHERE-ZIMPOL at the Very Large Telescope, which is specifically designed to search for the reflected, polarized visible light of giant exoplanets, is limited by reflection-induced, polarization-dependent beam shifts \citep{schmid2018sphere}.
Such shifts also affect interferometric polarization measurements with the SPeckle Polarimeter at the Sternberg Astronomical Institute 2.5-m telescope \citep{safonov2019differential}.
The beam shifts become apparent for these instruments due to the unprecedented polarimetric sensitivity and spatial resolution they achieve.

The polarization aberrations of an astronomical telescope and instrument can be numerically computed with polarization ray tracing \citep{breckinridge2015polarization}.
First, the paths of the rays of light are traced through the optical system using geometrical optics, but instead of the intensity, the electric field components of the rays are computed upon each reflection or transmission \citep[e.g.,][]{waluschka1989polarization, chipman1989polarization, yun2011three1, yun2011three2}.
Each point in the exit pupil is then associated with a Jones matrix.
In this way, the Jones pupil, which maps the changes in the electric fields between the entrance and exit pupils of the system, is calculated \citep{totzeck2005describe}.
Finally, the intensity in the focal plane (i.e.,~the PSF) is computed in the Fraunhofer approximation through spatial Fourier transforms over the Jones pupil.
Several studies have used polarization ray tracing to model the polarization aberrations of proposed and future high-contrast imagers and telescopes, such as the Roman Space Telescope \citep{krist2017wfirst}, HabEx \citep{davis2018habex, breckinridge2018terrestrial}, LUVOIR \citep{sabatke2018polarization, will2019effects}, PICTURE-C \citep{mendillo2019polarization}, and the three extremely large telescopes \citep{anche2018estimation, anche2023polarization}. 
However, these numerical computations give little insight into the full range of aberrations, their origin and characteristics, and the relative importance of amplitude and phase effects.

\citet{breckinridge2015polarization} use polarization ray tracing to analyze a three-mirror system consisting of a Cassegrain telescope followed by a flat fold mirror, and find two beam-shift effects that both originate from the oblique reflection off the flat mirror. 
The authors find phase gradients (i.e.,~wavefront tilts) in the Jones pupil that have opposite directions for the linearly polarized components parallel and perpendicular to the plane of incidence of the fold mirror.
In the focal plane, these gradients cause the orthogonally polarized components of the PSF to shift in opposite directions, thereby broadening the resulting PSF in intensity.
Furthermore, the authors find PSF components that couple the light from one orthogonal polarization into the other.
These PSF components, which they call ghost PSFs, have two peaks, one on either side of the plane of incidence.

Sub-wavelength, polarization-dependent shifts of a beam of light induced by reflection off a flat metallic mirror are also extensively described in the physics literature \citep[for overviews, see][]{aiello2008role, gotte2012generalized, bliokh2013goos}.
These shifts are referred to as the Goos-H\"anchen (GH) and Imbert-Federov (IF) shifts and occur in the directions parallel and perpendicular to the plane of incidence, respectively.
Both shifts are further divided into a spatial and an angular shift.
The spatial shifts are displacements of the entire beam of light upon reflection, and the angular shifts refer to angular deviations of the beam upon reflection. 
As such, the four shifts are considered first-order corrections to the laws of geometrical optics due to diffraction within a beam of light of finite width; the Fresnel equations only apply to infinitely extended interfaces, and a correct description of light reflected off an interface must therefore take into account the finite beam size.
The GH and IF shifts are derived from first principles through full diffraction calculations and are described using closed-form mathematical expressions specifying the centroid of the intensity of a reflected Gaussian beam \citep[e.g.,][]{aiello2007reflection, aiello2008role}.
All four shifts have been experimentally validated for metallic reflections \citep{merano2007observation, aiello2009duality, hermosa2011spin}.
\citet{schmid2018sphere} show in their analysis of the beam shifts of SPHERE-ZIMPOL that the spatial GH shift is likely the same as the shift arising from phase gradients in the Jones pupil as described by \citet{breckinridge2015polarization}.

In this paper, we aim to understand polarization aberrations produced by reflection off flat metallic mirrors at the fundamental level and seek to unify the two views of the beam shifts from polarization ray tracing and full diffraction calculations in the physics literature.
To this end, we determine the beam shifts from the polarization ray tracing of the reflection of a beam of light with a uniform (or top-hat) intensity profile (as applies to astronomical telescopes and instruments), and compare the resulting shifts to the spatial and angular GH and IF shifts as predicted by the closed-form expressions derived for Gaussian beams.
We investigate whether the GH and IF shifts are reproduced by polarization ray tracing or whether they are additional effects that we need to take into account for astronomical instruments.
In addition, we study the origin and characteristics of the shifts and determine how the size and direction of the shifts depend on the beam intensity profile, incident polarization state, angle of incidence, mirror material, and wavelength.
Finally, we examine how these shifts affect the performance of high-contrast imagers and how we can mitigate them in (future) diffraction-limited astronomical telescopes and instruments.

The outline of this paper is as follows.
In Sect.~\ref{sec:definitions} we describe the conventions and definitions of the mathematics used throughout the paper.
Subsequently, in Sect.~\ref{sec:numerical}, we outline the polarization ray tracing of the reflection of a beam of light off a flat metallic mirror and the determination of the beam shifts.
In Sect.~\ref{sec:beam_shifts} we then explain the origin and characteristics of the spatial and angular GH and IF shifts and their relation to shifts found using polarization ray tracing. 
We also show the dependence of the size and direction of the shifts on the incident polarization state and angle of incidence.
In Sect.~\ref{sec:discussion} we investigate the polarization structure in the PSF induced by the beam shifts and the effect of the beam shifts on polarimetric measurements.
In the same section we also examine the size of the beam shifts for various mirror materials and wavelengths, and discuss and refine the approaches to mitigate the beam shifts.
Finally, we show a table summarizing the properties of the four beam shifts at the end of Sect.~\ref{sec:discussion} and present conclusions in Sect.~\ref{sec:conclusions}.



%
%

\section{Conventions and definitions}
\label{sec:definitions}

In this section, we outline the conventions and definitions used throughout this paper.
In the literature, the mathematical definitions underlying the descriptions of polarization aberrations and beam shifts are often incomplete and not consistent among different studies. 
This can lead to errors in the physical interpretation, for example with the handedness of the circular polarization or the direction of the beam shifts. 
We therefore describe our definitions extensively and have carefully checked our equations for consistency.
As such, this paper provides a complete reference for the correct computation of the polarization aberrations and beam shifts.
To enable easy comparison of our results with those from the physics literature, we use the same definitions as \citet{aiello2007reflection}, \citet{merano2007observation}, \citet{aiello2008role}, \citet{aiello2009duality}, and \citet{hermosa2011spin}. 
For the description of the polarization of light, these definitions are consistent with the definitions adopted by the International Astronomical Union \citep[see e.g.,][]{hamaker1996understanding}.
We present the mathematics to describe light and its polarization in Sect.~\ref{sec:polarization} and discuss metallic reflection in Sect.~\ref{sec:reflection}.


\subsection{Polarization of light}
\label{sec:polarization}

\begin{figure}[!t]
\centering 
\includegraphics[width=\hsize]{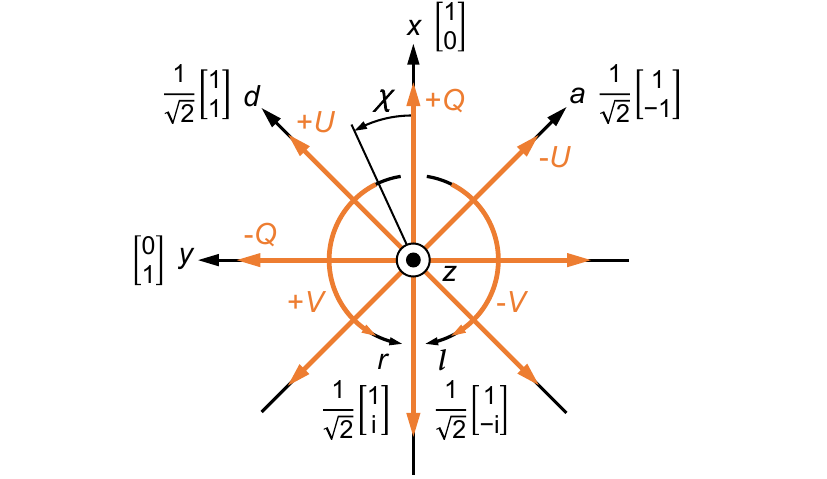} 
\caption{Definition of the three reference frames (or bases) and the Stokes parameters to describe the electric field components and polarization of an electromagnetic wave. The light propagates along the $z$-axis out of the paper toward the reader. In the $xyz$-basis, the $x$-axis ($y$-axis) is oriented in the vertical (horizontal) direction. In the $daz$-basis, the $d$-axis ($a$-axis) is oriented in the diagonal (antidiagonal) direction, at $45\degr$ counterclockwise (clockwise) from the $x$-axis. In the $rlz$-basis, $r$ and $l$ represent the right-handed and left-handed circularly polarized components. For each reference frame, the basis Jones vectors, expressed in the $xyz$-bases, are indicated. The Stokes parameters are shown in orange with the plus sign (minus sign) indicating that the Stokes parameter is positive (negative) in that direction. The angle of linear polarization $\chi$ is defined positive for a counterclockwise rotation from the $x$-axis.} 
\label{fig:reference_frame} 
\end{figure}

We shall consider a monochromatic, polarized light wave propagating in the positive $z$-direction of a Cartesian reference frame (or basis) $xyz$ as shown in Fig.~\ref{fig:reference_frame}. 
The transverse electric field components of this light wave in the vertical $x$- and horizontal $y$-directions can then be described as follows \citep[see e.g.,][]{born2013principles}:
\begin{align} 
\tilde{E}_x(z,t) &= A_x\cos\left(k z - \omega t + \varphi_x\right) = \operatorname{Re}\left[A_x\mathrm{e}^{\mathrm{i}\varphi_x}\mathrm{e}^{\mathrm{i}(k z - \omega t)}\right], \label{eq:light_wave_x} \\[0.2cm]
\tilde{E}_y(z,t) &= A_y\cos\left(k z - \omega t + \varphi_y\right) = \operatorname{Re}\left[A_y\mathrm{e}^{\mathrm{i}\varphi_y}\mathrm{e}^{\mathrm{i}(k z - \omega t)}\right], \label{eq:light_wave_y}
\end{align}  
where $t$ is time, $\omega > 0$ is the angular frequency, $k = 2\uppi/\lambda$ is the wave number with $\lambda$ the wavelength, $A_x$ and $A_y$ are the amplitudes, $\varphi_x$ and $\varphi_y$ are the initial phases, $\operatorname{Re}[\dots]$ denotes the real part, and $\mathrm{i}$ is the imaginary unit. 
On the right side of Eqs.~(\ref{eq:light_wave_x}) and (\ref{eq:light_wave_y}), the factor $\exp{[\mathrm{i}(k z - \omega t)]}$ only describes the propagation of the light wave.
The polarization of the wave can therefore be described by a Jones vector $\boldsymbol{E}$:
\begin{align} 
\boldsymbol{E} = \begin{bmatrix} E_x \\ E_y \end{bmatrix} = \begin{bmatrix} A_x\mathrm{e}^{\mathrm{i}\varphi_x} \\ A_y\mathrm{e}^{\mathrm{i}\varphi_y} \end{bmatrix},
\label{eq:jones_vector} 	 
\end{align}  
where $E_x$ and $E_y$ are the complex electric field components.
%

As an alternative way to describe the polarization, we can define a set of Stokes parameters (see Fig.~\ref{fig:reference_frame}):
\begin{alignat}{4}
I &= E_x E_x^* + E_y E_y^* &&= A_x^2 + A_y^2 &&= I_x + I_y = I_d + I_a \nonumber \\[0.2cm]
  &                        &&                &&= I_r + I_l = 1, \label{eq:stokes_i} \\[0.2cm]
Q &= E_x E_x^* - E_y E_y^* &&= A_x^2 - A_y^2 &&= I_x - I_y, \label{eq:stokes_q} \\[0.2cm]
U &= E_x E_y^* + E_y E_x^* &&= 2 A_x A_y \cos\delta &&= I_d - I_a, \label{eq:stokes_u} \\[0.2cm]
V &= \mathrm{i}\left(E_x E_y^*- E_y E_x^* \right) &&= 2 A_x A_y \sin\delta &&= I_r - I_l, \label{eq:stokes_v}
\end{alignat}
where the asterisk denotes the complex conjugate, $\delta = \varphi_y - \varphi_x$ is the phase difference between the $y$- and $x$-components of the electric field, and $I_x$ and $I_y$ are the intensities of the $x$- and $y$-components of the electric field. 
The variables $I_d$ and $I_a$ are the intensities of the $d$- and $a$-components in the basis of the diagonal and antidiagonal polarizations, $daz$, and $I_r$ and $I_l$ are the intensities of the $r$- and $l$-components in the basis of the right-handed and left-handed circular polarizations, $rlz$ (see Fig.~\ref{fig:reference_frame}).
Stokes $I$ is the total intensity, positive (negative) Stokes $Q$ describes linear polarization in the vertical $x$-direction (horizontal $y$-direction), positive (negative) Stokes $U$ describes linear polarization in the diagonal (antidiagonal) direction, $45\degr$ counterclockwise (clockwise) from the $x$-direction, and positive (negative) Stokes $V$ describes right-handed (left-handed) circular polarization. 
Whereas the $xyz$-basis is the natural basis of Stokes $Q$, the $daz$- and $rlz$-bases are the natural bases of Stokes $U$ and $V$, respectively.
Because we normalize the total intensity, that is, we set $I = 1$ in Eq.~(\ref{eq:stokes_i}), $Q$, $U$, and $V$ have values between $1$ and $-1$. 
We note that Eqs.~(\ref{eq:stokes_i})--(\ref{eq:stokes_v}) are strictly speaking only valid for 100\% polarized, monochromatic light. However, for quasi-monochromatic light, whether 100\% polarized, partially polarized, or unpolarized, we simply need to take the time averages over the terms in the equations. 

From Eqs.~(\ref{eq:stokes_i}) and (\ref{eq:stokes_q}), we can derive expressions for the intensities of the $x$- and $y$-components of the electric field: 
\begin{align}
I_x = \dfrac{1 + Q}{2}, \label{eq:i_x} \\[0.2cm]
I_y = \dfrac{1 - Q}{2}. \label{eq:i_y}
\end{align}
%
Although these two equations are simple, they are important, and we use them in all closed-form expressions for the beam shifts in Sect.~\ref{sec:beam_shifts}. Finally, we assemble the Stokes parameters in a Stokes vector $\boldsymbol{S}$:
\begin{align} 
\boldsymbol{S} = \begin{bmatrix} I \\ Q \\ U \\ V \end{bmatrix},
\label{eq:stokes_vector} 	 
\end{align}  
and define the degree of linear polarization $P$ (which for $I = 1$ is equal to the linearly polarized intensity) and angle of linear polarization $\chi$ (see Fig.~\ref{fig:reference_frame}) as follows:
\begin{align} 
P &= \sqrt{Q^2 + U^2}, \label{eq:dolp} \\[0.2cm]
\chi &= \dfrac{1}{2}\arctan\left(\dfrac{U}{Q}\right). \label{eq:aolp}
\end{align}  
%


\subsection{Metallic reflection}
\label{sec:reflection}

\begin{figure}[!t]
\centering
\includegraphics[width=\hsize]{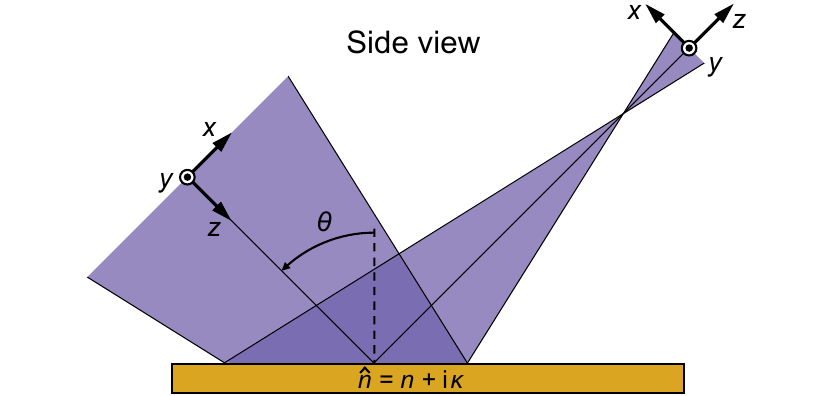} 
\caption{Schematic of the reflection of a beam of light off a flat metallic mirror with complex refractive index $\hat{n} = n + \mathrm{i}\kappa$. The central ray of the beam hits the mirror at an angle of incidence $\theta$ measured with respect to the normal to the surface of the mirror. The orientation of the $xyz$ reference frame before and after reflection is indicated.} 
\label{fig:reflection} 
\end{figure}

Using this mathematically consistent description of light and its polarization, we can describe the reflection of light using the Fresnel equations in the geometric polarization ray-tracing approximation.
We shall consider the central ray of a beam of light incident on a flat metallic mirror as shown in Fig.~\ref{fig:reflection}.
Describing this ray as a plane electromagnetic wave, we decompose the incident electric field into the $p$- and $s$-polarized components that are parallel and perpendicular to the plane of incidence, respectively. 
For this central ray, the $p$- and $s$-directions correspond to the $x$- and $y$-directions, respectively.
Assuming the refractive index of the incident medium (air) to be equal to 1, we compute the complex Fresnel reflection coefficients $r_p$ and $r_s$ as follows \citep[see e.g.,][]{born2013principles}:
\begin{alignat}{2}
r_p &= \dfrac{\hat{n}^2\cos\theta - \sqrt{\hat{n}^2 - \sin^2\theta}}{\hat{n}^2\cos\theta + \sqrt{\hat{n}^2 - \sin^2\theta}} &&= R_p\mathrm{e}^{\mathrm{i}\phi_p}, \label{eq:fresnel_rp} \\[0.2cm]
r_s &= \dfrac{\cos\theta - \sqrt{\hat{n}^2 - \sin^2\theta}}{\cos\theta + \sqrt{\hat{n}^2 - \sin^2\theta}} &&= R_s\mathrm{e}^{\mathrm{i}\phi_s}, \label{eq:fresnel_rs}
\end{alignat}  
where $\theta$ is the central angle of incidence (see Fig.~\ref{fig:reflection}) and $\hat{n} = n + \mathrm{i}\kappa$ is the complex refractive index of the mirror material, with $n$ and $\kappa$ the real and complex parts, respectively. The amplitudes $R_{p/s} = \lvert r_{p/s} \rvert$ specify the ratios of the amplitudes of the reflected and incident electric fields, while the phases $\phi_{p/s} = \arg{(r_{p/s})}$ describe the phase shifts between the reflected and incident electric fields.
%

Two important quantities related to the reflection coefficients are the diattenuation and the retardance, which can be considered to be the zeroth-order polarization aberrations. The diattenuation $\epsilon$ is defined as follows: 
\begin{align} 
\epsilon = \dfrac{R_s^2 - R_p^2}{R_s^2 + R_p^2},
\label{eq:diattenuation} 	 
\end{align}  
which ideally equals 0. When unpolarized light is incident on the mirror, a nonzero value of the diattenuation quantifies the amount of linearly polarized light that is created, that is, the instrumental polarization. The retardance $\varDelta$ is defined as follows:
\begin{align} 
\varDelta = \phi_s - \phi_p,
\label{eq:retardance} 	 
\end{align}  
which ideally equals $180\degr$. The latter value comes from the requirement that the electromagnetic wave before and after reflection is described by a right-handed triplet in terms of the electric field, the magnetic field, and the wave vector. For values other than $180\degr$, retardance results in the conversion of incident linearly polarized light into circularly polarized light and vice versa, that is, it produces polarimetric crosstalk.

\begin{figure}[!t]
\centering
\includegraphics[width=\hsize]{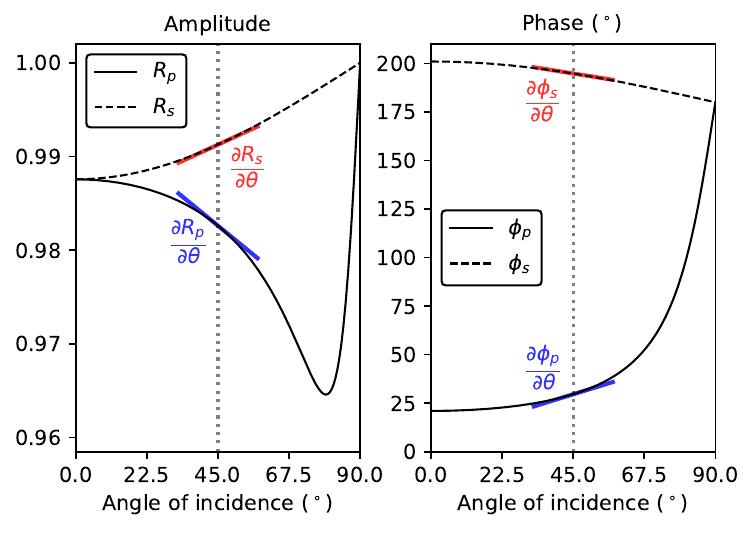} 
\caption{Amplitude (\emph{left}) and phase (\emph{right}) of the Fresnel reflection coefficients in the $p$- and $s$-directions as a function of the angle of incidence for gold with $\hat{n} = 0.188 + \mathrm{i}5.39$ at a wavelength of 820~nm.
The gradients in the amplitude and phase for an angle of incidence of $45\degr$ are indicated in blue for the $p$-direction and in red for the $s$-direction.} 
\label{fig:fresnel} 
\end{figure}

The physics of the beam shifts as described in Sect.~\ref{sec:beam_shifts} depends on the diattenuation and retardance as well as on the gradients of the amplitude and phase of the reflection coefficients with the angle of incidence.
Figure~\ref{fig:fresnel} shows the amplitude and phase of the reflection coefficients as a function of the angle of incidence for gold with $\hat{n} = 0.188 + \mathrm{i}5.39$ at a wavelength of 820~nm, corresponding to the configuration studied in \mbox{Sects.~\ref{sec:numerical}--\ref{sec:discussion}}.
From Fig.~\ref{fig:fresnel}~(left) it follows that the diattenuation, which is roughly the difference between the curves of $R_s$ and $R_p$ (see Eq.~(\ref{eq:diattenuation})), is zero at $\theta = 0\degr$, increases with increasing angle of incidence until it reaches a maximum around $\theta = 80\degr$, and then decreases again to zero at $\theta = 90\degr$.
In Fig.~\ref{fig:fresnel}~(right) we see that the retardance, which is the difference between the curves of $\phi_s$ and $\phi_p$ (see Eq.~(\ref{eq:retardance})), is $180\degr$ at $\theta = 0\degr$ and remains close to this value for small values of $\theta$.
For large $\theta$, the retardance decreases rapidly to $0\degr$ at $\theta = 90\degr$.
Fig.~\ref{fig:fresnel}~(left and right) also show the gradients in amplitude and phase at $\theta = 45\degr$ (similar to the phase gradients shown by \citealt{breckinridge2015polarization}).
Whereas the amplitude gradient $\partial R_s / \partial\theta$ is always positive for $\theta > 0\degr$, $\partial R_p / \partial\theta$ is initially negative, then becomes zero, and finally is positive for very large angles of incidence.
Lastly, for $\theta > 0\degr$ the phase gradients $\partial\phi_s / \partial\theta$ and $\partial\phi_p / \partial\theta$ are negative and positive, respectively, and monotonically decrease and increase with increasing angle of incidence.
%
%

%
%

\section{Beam shifts from polarization ray tracing}
\label{sec:numerical}

In this section, we describe the polarization ray tracing of a beam of light that reflects off a (flat) metallic mirror, following the methodology outlined in \citet{breckinridge2015polarization}, and the determination of the beam shifts that result. 
In Sect.~\ref{sec:beam_shifts} we compare the resulting shifts for various incident polarization states and angles of incidence to the predicted spatial and angular GH and IF shifts as derived for Gaussian beams.
We determine the centroid shifts of both the focal-plane intensity (i.e.,~the PSF) and the intensity in the exit-pupil plane because these planes are where the spatial shifts (shifts of the complete beam) and angular shifts (angular deviations as measured from the focus) should be visible. 
To enable a direct comparison of our results with the experimental measurements of the GH and IF shifts by \citet{merano2007observation}, \citet{aiello2009duality}, and \citet{hermosa2011spin}, we consider a (practically) identical configuration to the one used in those studies: a converging, monochromatic beam of light with an f-number of 61.3 that reflects off a flat gold mirror at a wavelength of 820~nm and with a focal distance of 11.9~cm.
Our configuration differs in that the beam of light is not Gaussian but has a uniform (or top-hat) intensity profile across the entrance pupil as is the case for astronomical telescopes and instruments.

\begin{figure*}[!htbp]
\centering
\subfloat{\includegraphics[trim=-3.5pt 38pt 0pt 0pt, clip, width=15.8cm]{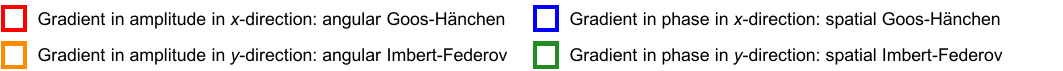}} \\
\subfloat{\includegraphics[trim=-2.5pt 5pt 5pt 3pt, clip, width=15.93cm]{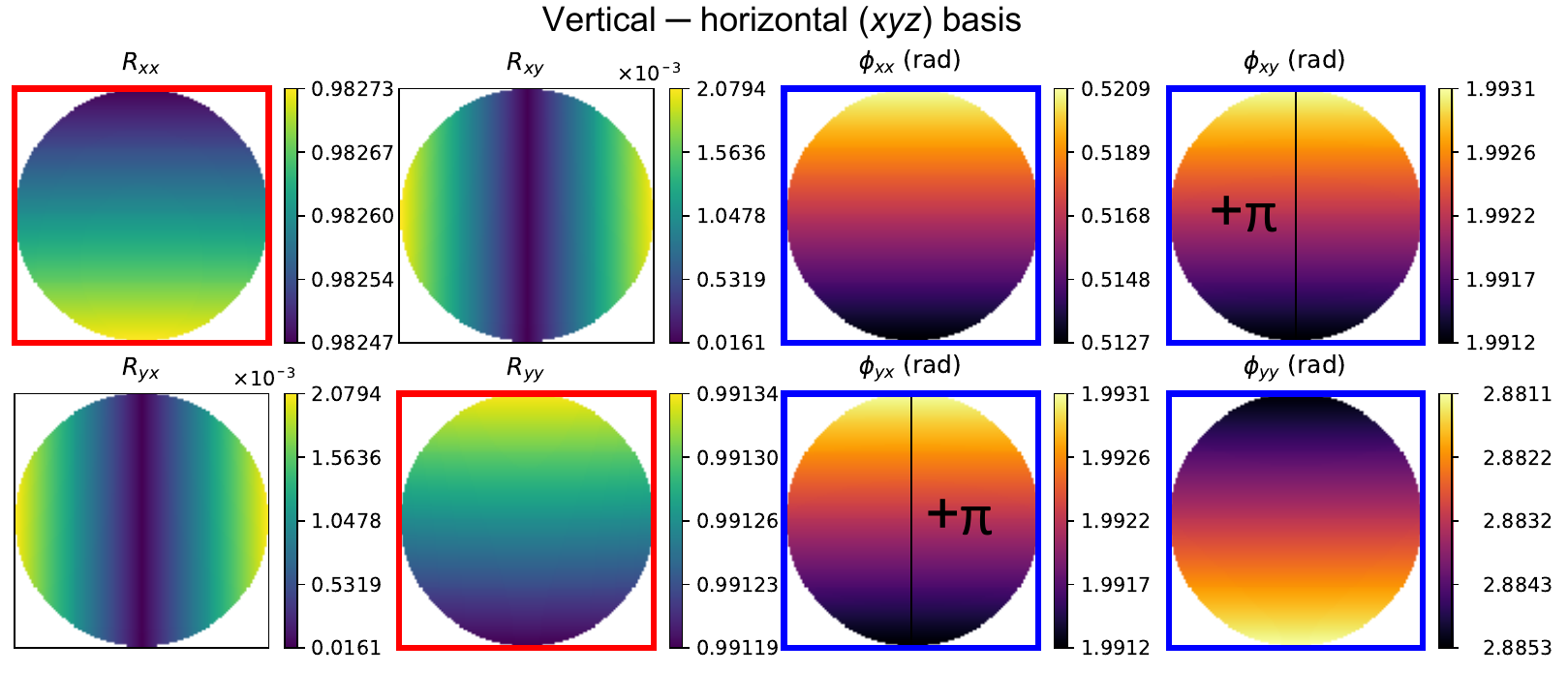}} \\
\subfloat{\includegraphics[trim=0pt 5pt 0pt 4pt, clip, width=15.8cm]{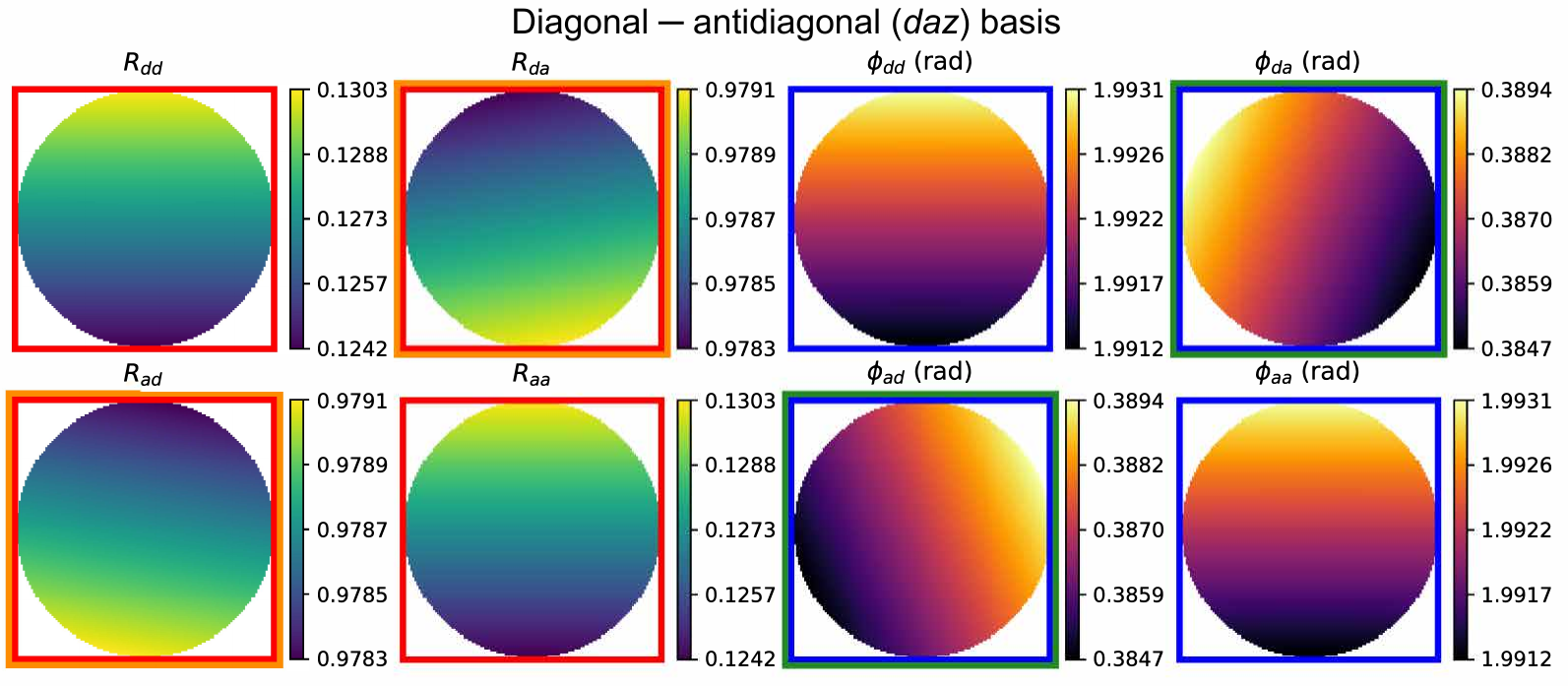}} \\
\subfloat{\includegraphics[trim=0pt 5pt 0pt 4pt, clip, width=15.8cm]{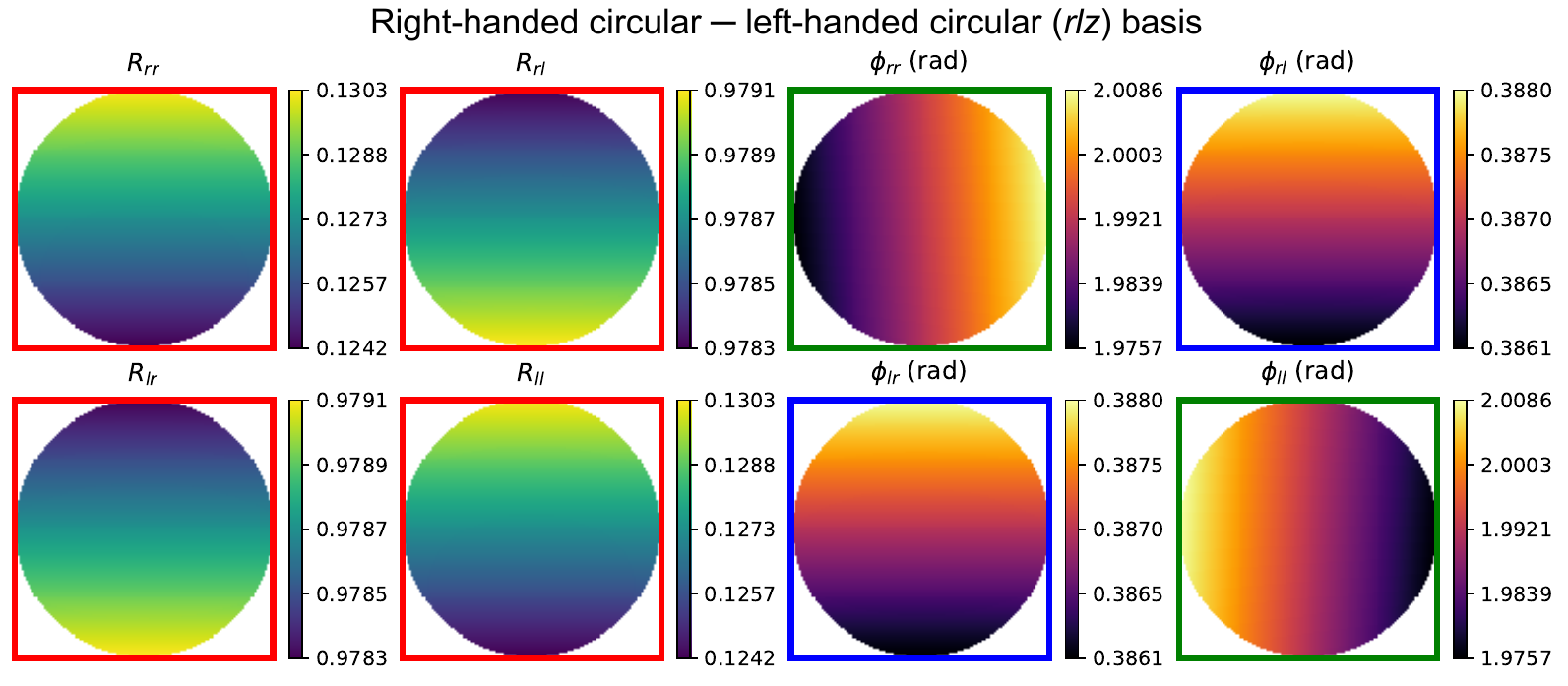}}
\caption{Jones pupil expressed in the $xyz$- (\emph{top}), $daz$- (\emph{center}), and $rlz$-bases (\emph{bottom}) at a wavelength of 820~nm for a converging beam of light with an f-number of 61.3 that reflects off gold at an angle of incidence of $45\degr$.
The panels in the first and second (third and fourth) columns show the amplitude (phase) of the Jones-pupil elements.
The positive $x$- and $y$-directions are upward and to the left, respectively. 
The values of the color maps are different among the panels.
The red, orange, blue, and green borders around the panels indicate the gradients that are visible and the specific beam shifts that these gradients cause (see the legend above the top panels).
The panels of $R_{da}$, $R_{ad}$, $\phi_{da}$, and $\phi_{ad}$ have two colored borders because these panels show a combination of two gradients.
To reveal the gradient in the panels of $\phi_{xy}$ and $\phi_{yx}$, $\uppi$ has been added to the phase in the left and right halves of the pupil, respectively.
}
\label{fig:jones_pupil} 
\end{figure*}

As the first step in our analysis, we compute the Jones pupil that describes the electric-field response in the exit pupil upon reflection.
We only describe this computation briefly here \citep[for detailed descriptions see e.g.,][]{waluschka1989polarization, gotte2012generalized}.
We use the definitions as shown in Fig.~\ref{fig:reflection} and decompose the beam of light into a set of rays that each can be described by a plane electromagnetic wave.
For each ray, we compute the angle of incidence and, using Eqs.~(\ref{eq:fresnel_rp}) and (\ref{eq:fresnel_rs}), the corresponding Fresnel reflection coefficients in the local $p$- and $s$-directions.
Subsequently, we calculate the orientation of the local plane of incidence for each ray.
Finally, we compute the Jones pupil as the set of Jones matrices describing the reflection of each ray, taking into account the orientation of the local plane of incidence and the change of sign of the $x$-coordinate of the ray upon reflection.
The resulting Jones pupil $J_{xyz}$, which is expressed in the $xyz$-basis, can be written as follows:
\begin{align} 
J_{xyz} = \begin{bmatrix} J_{xx} & J_{xy} \\ J_{yx} & J_{yy} \end{bmatrix} = \begin{bmatrix} R_{xx}\mathrm{e}^{\mathrm{i}\phi_{xx}} & R_{xy}\mathrm{e}^{\mathrm{i}\phi_{xy}} \\ R_{yx}\mathrm{e}^{\mathrm{i}\phi_{yx}} & R_{yy}\mathrm{e}^{\mathrm{i}\phi_{yy}} \end{bmatrix},
\label{eq:jones_pupil_xy} 	 
\end{align}  
where $J_{xx}$ to $J_{yy}$ are the complex Jones-pupil elements describing the contribution of the $x$- or $y$-polarized components of the incident electric field (in the entrance pupil) to the $x$- or $y$-polarized components of the reflected electric field (in the exit pupil).
The amplitudes and phases of the Jones-pupil elements, which define the ratios of the amplitudes and the phase shifts of the reflected and incident electric fields, are denoted $R_{xx}$ to $R_{yy}$ and $\phi_{xx}$ to $\phi_{yy}$, respectively.
The Jones pupil $J_{xyz}$ for reflection with an angle of incidence of $45\degr$ is shown in Fig.~\ref{fig:jones_pupil} (top).
%
%
%
\begin{figure*}[!htbp]
\centering
\includegraphics[width=18.43cm]{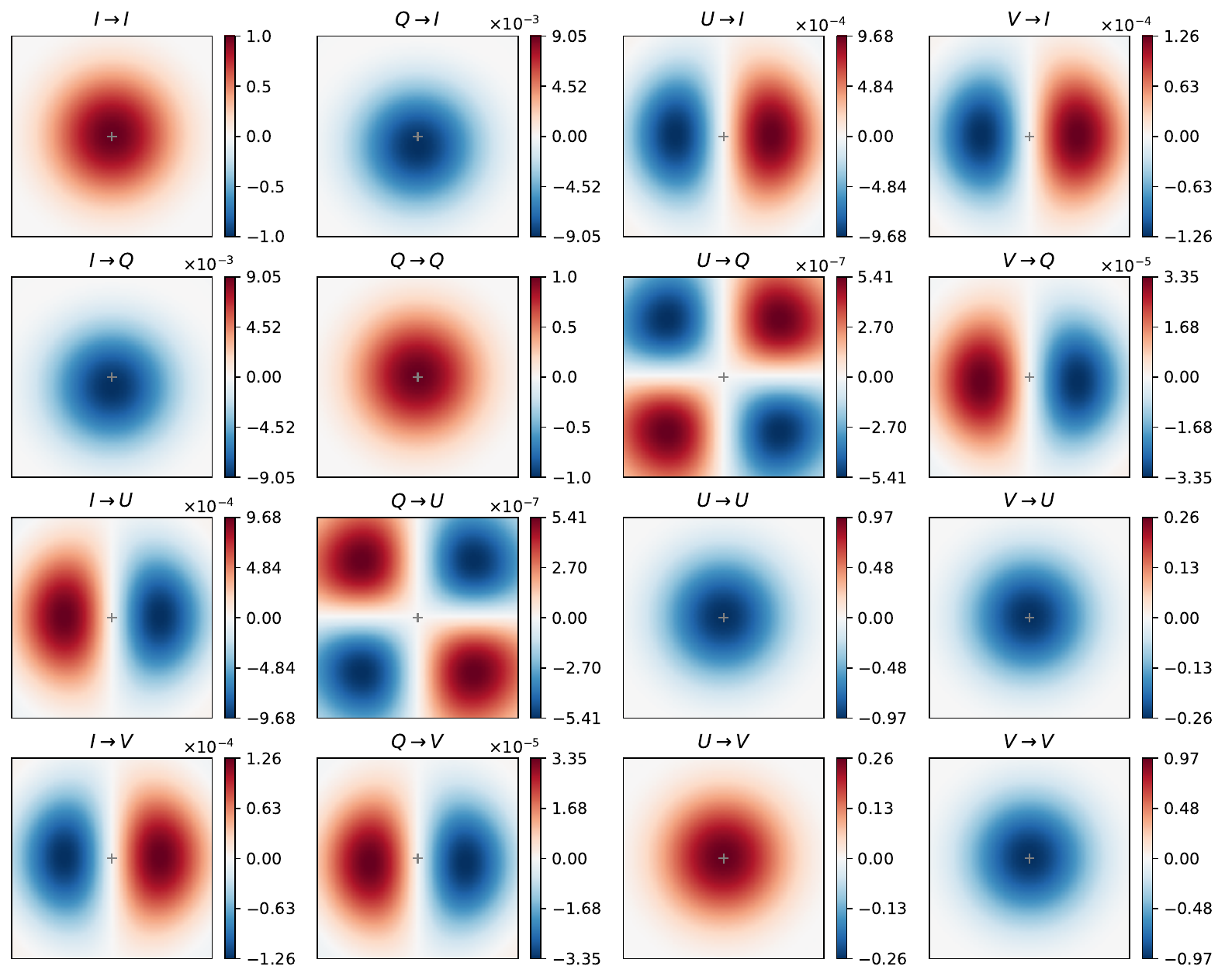} 
\caption{Point-spread matrix (PSM) at a wavelength of 820~nm for a converging beam of light with an f-number of 61.3 that reflects off gold at an angle of incidence of $45\degr$. The panels show the central $100~\upmu$m $\times$ $100~\upmu$m of the PSM elements. The positive $x$- and $y$-directions are upward and to the left, respectively. The gray plus signs indicate the centroids of the PSM elements in the absence of diffraction and aberrations. The values of the color maps are different among the panels.}
\label{fig:psm} 
\end{figure*} 
%

The Jones pupil is a crucial ingredient for our understanding of the beam shifts in Sect.~\ref{sec:beam_shifts}.
In that context, it is useful to also express the Jones pupil in the basis of the diagonal and antidiagonal polarizations, $daz$, and the basis of the right-handed and left-handed circular polarizations, $rlz$, as defined in Fig.~\ref{fig:reference_frame}.
The Jones pupils in the $daz$- and $rlz$-bases, $J_{daz}$ and $J_{rlz}$, are defined as follows:
\begin{alignat}{2} 
J_{daz} &= T_{daz} J_{xyz} T_{daz}^{-1} &&= \begin{bmatrix} R_{dd}\mathrm{e}^{\mathrm{i}\phi_{dd}} & R_{da}\mathrm{e}^{\mathrm{i}\phi_{da}} \\ R_{ad}\mathrm{e}^{\mathrm{i}\phi_{ad}} & R_{aa}\mathrm{e}^{\mathrm{i}\phi_{aa}} \end{bmatrix}, \label{eq:jones_pupil_da} \\[0.2cm]
J_{rlz} &= T_{rlz} J_{xyz} T_{rlz}^{-1} &&= \begin{bmatrix} R_{rr}\mathrm{e}^{\mathrm{i}\phi_{rr}} & ~R_{rl}\mathrm{e}^{\mathrm{i}\phi_{rl}} \\ R_{lr}\mathrm{e}^{\mathrm{i}\phi_{lr}} & ~R_{ll}\mathrm{e}^{\mathrm{i}\phi_{ll}} \end{bmatrix}, \label{eq:jones_pupil_rl}
\end{alignat}  
where $R_{dd}$ to $R_{ll}$ and $\phi_{dd}$ to $\phi_{ll}$ are the amplitudes and phases of the Jones-pupil elements and $^{-1}$ denotes the inverse of a matrix.
The matrices $T_{daz}$ and $T_{rlz}$ describe the transformations from the $xyz$-basis to the $daz$- and $rlz$-bases, respectively, and are given by:
\begin{align} 
T_{daz} &= \dfrac{1}{\sqrt{2}} \begin{bmatrix} 1 & 1 \\ 1 & -1 \end{bmatrix}, \label{eq:transform_da} \\[0.2cm]
T_{rlz} &= \dfrac{1}{\sqrt{2}} \begin{bmatrix} 1 & -\mathrm{i} \\ 1 & \mathrm{i} \end{bmatrix}. \label{eq:transform_rl}
\end{align}  
The Jones pupils $J_{daz}$ and $J_{rlz}$ for reflection with an angle of incidence of $45\degr$ are shown in Fig.~\ref{fig:jones_pupil} (center) and Fig.~\ref{fig:jones_pupil} (bottom), respectively.

As the next step, we compute the amplitude-response matrix (ARM) specifying the electric-field response in the focal plane (expressed in the $xyz$-basis). The ARM is computed as follows:
\begin{align} 
\mathit{ARM} = \begin{bmatrix} \mathscr{F}(J_{xx}) & \mathscr{F}(J_{xy}) \\ \mathscr{F}(J_{yx}) & \mathscr{F}(J_{yy}) \end{bmatrix} = \begin{bmatrix} R'_{xx}\mathrm{e}^{\mathrm{i}\phi'_{xx}} & R'_{xy}\mathrm{e}^{\mathrm{i}\phi'_{xy}} \\ R'_{yx}\mathrm{e}^{\mathrm{i}\phi'_{yx}} & R'_{yy}\mathrm{e}^{\mathrm{i}\phi'_{yy}} \end{bmatrix},
\label{eq:arm} 	 
\end{align}  
where $\mathscr{F}(\dots)$ denotes the spatial Fourier transform over a Jones-pupil element and $R'_{xx}$ to $R'_{yy}$ and $\phi'_{xx}$ to $\phi'_{yy}$ denote the amplitudes and phases, respectively, of the ARM elements. By using the spatial Fourier transform for the computation of the ARM we assume that the Fraunhofer approximation to diffraction applies, which is the case for beams with absolute f-numbers larger than ${\sim}5$ \citep[see e.g.,][]{mcguire1990diffraction}. The ARM for reflection with an angle of incidence of $45\degr$ is shown in Fig.~\ref{fig:arm}.

Next, we calculate the point-spread matrix (PSM), which is the Mueller-matrix representation of the PSF and describes the intensity response in the focal plane for any incident Stokes vector, whether 100\% polarized, partially polarized, or unpolarized.
The PSM is calculated as follows:
\begin{align} 
\mathit{PSM} = C(\mathit{ARM} \otimes \mathit{ARM}^{*})C^{-1}
\label{eq:psm_calc} 	 
\end{align}  
where $\otimes$ denotes the Kronecker product, the asterisk indicates the element-wise complex conjugate, and the matrix $C$ is given by \citep[see e.g.,][]{espinosa2008transformation}:
\begin{align} 
C = \begin{bmatrix} 1 & 0 & 0 & 1 \\ 1 & 0 & 0 & -1 \\ 0 & 1 & 1 & 0 \\ 0 & \mathrm{i} & -\mathrm{i} & 0 \end{bmatrix}.
\label{eq:psm_submatrix} 	 
\end{align}  
The PSM can be written as follows:
\begin{align} 
\mathit{PSM}= \begin{bmatrix*}[l] 
I \hspace{-2pt}\rightarrow\hspace{-2pt} I & Q \hspace{-2pt}\rightarrow\hspace{-2pt} I & U \hspace{-2pt}\rightarrow\hspace{-2pt} I & V \hspace{-2pt}\rightarrow\hspace{-2pt} I \\
I \hspace{-2pt}\rightarrow\hspace{-2pt} Q & Q \hspace{-2pt}\rightarrow\hspace{-2pt} Q & U \hspace{-2pt}\rightarrow\hspace{-2pt} Q & V \hspace{-2pt}\rightarrow\hspace{-2pt} Q \\
I \hspace{-2pt}\rightarrow\hspace{-2pt} U & Q \hspace{-2pt}\rightarrow\hspace{-2pt} U & U \hspace{-2pt}\rightarrow\hspace{-2pt} U & V \hspace{-2pt}\rightarrow\hspace{-2pt} U \\
I \hspace{-2pt}\rightarrow\hspace{-2pt} V & Q \hspace{-2pt}\rightarrow\hspace{-2pt} V & U \hspace{-2pt}\rightarrow\hspace{-2pt} V & V \hspace{-2pt}\rightarrow\hspace{-2pt} V \end{bmatrix*},
\label{eq:psm} 	 
\end{align}
where each element $A \hspace{-2pt}\rightarrow\hspace{-2pt} B$ describes the contribution of the incident Stokes parameter $A$ to the resulting Stokes parameter $B$. The PSM for reflection with an angle of incidence of $45\degr$ is shown in Fig.~\ref{fig:psm}. We note that the same PSM can also be obtained by computing the ARM (Eq.~(\ref{eq:arm})) from the Jones pupil expressed in the $daz$- or $rlz$-bases and replacing the matrix $C$ in Eqs.~(\ref{eq:psm_calc}) and (\ref{eq:psm_submatrix}) with the appropriate matrix corresponding to those bases.

As the final step, we determine the beam shifts in the exit pupil and the focal plane.
To this end, we define an incident Jones vector or Stokes vector with a uniform intensity profile and polarization state.
For the determination of the shift in the exit pupil, we right-multiply the Jones pupil by the incident Jones vector to obtain the Jones vector in the pupil plane.
Subsequently, we compute the intensity distribution in the pupil plane as the sum of squares of the amplitudes of the latter Jones vector.
Finally, we calculate the beam shift as the offset of the centroid of the intensity distribution with respect to the beam position in the absence of diffraction and aberrations.
To determine the beam shift in the focal plane, we compute the Stokes vector after reflection by right-multiplying the PSM by the incident Stokes vector.
We then retrieve the intensity image from the first element of the resulting Stokes vector and determine the shift as the offset of the centroid with respect to the beam position in the absence of diffraction and aberrations. 

%
%

\section{Explanation of beam shifts and comparison to polarization ray tracing}
\label{sec:beam_shifts}

In this section, we explain the spatial and angular GH and IF shifts and compare them to the shifts found using polarization ray tracing.
We analytically describe the four shifts using the closed-form expressions from \citet{aiello2008role}. 
These expressions are derived \citep[see][]{aiello2007reflection} by decomposing an incident, uniformly polarized Gaussian beam of light into the angular spectrum of plane waves \citep[e.g.,][]{born2013principles} and computing the effect of the reflection on each wave. Because the plane waves are infinitely extended, the Fresnel equations can be applied without making any approximations. 
The decomposition into plane waves is equivalent to a Fourier transform of the electric field at the mirror interface.
The resulting reflected plane waves are then integrated over, and the shift is calculated as the shift of the centroid of the intensity of the beam.
The expressions depend on the Fresnel reflection coefficients at the central angle of incidence and the complex electric-field components of the incident beam.
We have rewritten the expressions in terms of the more familiar Stokes parameters to make the expressions easier to understand and enable the computation of the shifts for any incident polarization state.

For each of the four shifts, which generally occur simultaneously, we explain the origin and characteristics, and analytically compute the size and direction as a function of the angle of incidence for different incident polarization states.
We consider 100\% linearly polarized light with angles of linear polarization $\chi$ ranging from $0\degr$ to $180\degr$ in steps of $22.5\degr$, 100\% right-handed and left-handed circularly polarized light (i.e.,~$V = 1$ and $V = -1$, respectively), and unpolarized light.
For these same polarization states, we numerically compute the shifts from the polarization ray tracing as outlined in Sect.~\ref{sec:numerical} and compare the results to the analytical computations. 
We also explain the shifts using the Jones pupil and the PSM.
We discuss the spatial and angular GH shifts in Sects.~\ref{sec:sgh} and \ref{sec:agh} and the spatial and angular IF shifts in Sects.~\ref{sec:sif} and \ref{sec:aif}.
For easy reference, an overview of the properties of the four beam shifts is shown in Table~\ref{tab:beam_shifts} of Sect.~\ref{sec:overview_beam_shifts}.


\subsection{Spatial Goos-H\"anchen shift}
\label{sec:sgh}

\begin{figure}[!t]
\centering
\includegraphics[trim=0pt -10pt 0pt 0pt, clip, width=\hsize]{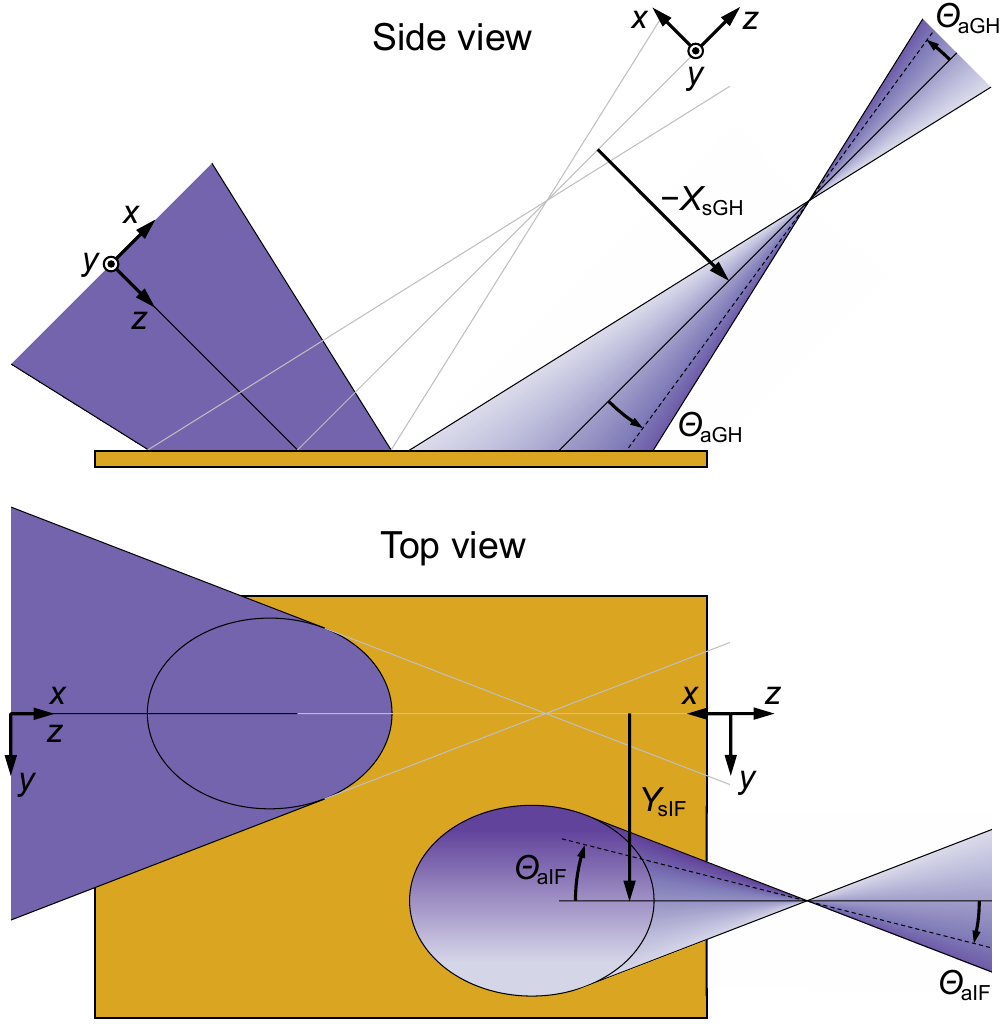}
\caption{Schematic showing the definitions of the spatial and angular GH shifts, $X_\mathrm{sGH}$ and $\varTheta_\mathrm{aGH}$ (\emph{top}), and the spatial and angular IF shifts, $Y_\mathrm{sIF}$ and $\varTheta_\mathrm{aIF}$ (\emph{bottom}), for an (initially converging) beam of light incident on a metallic mirror. Darker colors within the reflected beam indicate a higher relative intensity. The orientation of the $xyz$ reference frame before and after reflection is indicated. Positive spatial GH and IF shifts are directed in the positive $x$- and $y$-directions, respectively, after reflection (the spatial GH shift is shown in the negative direction). The angular GH and IF shifts are positive for a right-handed rotation around the $y$-axis and a left-handed rotation around the $x$-axis, respectively. For clarity the size of the shifts is shown extremely exaggerated.} 
\label{fig:beam_shifts} 
\end{figure}

The spatial GH shift, $X_\mathrm{sGH}$, is a displacement of the entire beam of light upon reflection and occurs in the plane of incidence \citep[e.g.,][]{goos1947neuer, merano2007observation, aiello2008role, aiello2009duality, gotte2012generalized, bliokh2013goos}. 
Figure~\ref{fig:beam_shifts} (top) shows a schematic with the definition of the spatial GH shift.
The shift is independent of the divergence angle of the incident beam (i.e.,~the f-number) and does not depend on whether the reflection occurs in the focus or the converging or diverging parts of the beam.
From the perspective of the plane-wave decomposition, the spatial GH shift can be understood from a 2D picture of the beam of light, looking from a direction perpendicular to the plane of incidence (i.e.,~the side view as shown in Fig.~\ref{fig:beam_shifts}, top).
Each plane wave of the beam has a different angle of incidence and therefore acquires a correspondingly different phase shift upon reflection.
This results in a gradient in phase over the range of angles of incidence (see Fig.~\ref{fig:fresnel}, right).
Integrating over all reflected plane waves, this then results in a shift of the entire beam parallel to the plane of incidence. The integration is equivalent to an inverse Fourier transform, which explains how a phase gradient is equivalent to a shift of the entire beam on the mirror.
%

The spatial GH shift can be computed as follows: 
\begin{align} 
X_\mathrm{sGH} = \dfrac{\lambda}{2 \uppi} \dfrac{\dfrac{\partial \phi_p}{\partial \theta} R_p^2 I_x + \dfrac{\partial \phi_s}{\partial \theta} R_s^2 I_y}{R_p^2 I_x + R_s^2 I_y},
\label{eq:sgh} 	 
\end{align}  
where $R_p$ and $R_s$ (from Eqs.~(\ref{eq:fresnel_rp}) and (\ref{eq:fresnel_rs})) and the phase gradients $\partial\phi_p / \partial\theta$ and $\partial\phi_s / \partial\theta$ (see Fig.~\ref{fig:fresnel}, right) are computed at the central angle of incidence of the beam, and $I_x$ and $I_y$ are the intensities of the components of the light polarized in the $x$- and $y$-direction, respectively. 
These intensities only depend on the incident Stokes $Q$ and follow from Eqs.~(\ref{eq:i_x}) and (\ref{eq:i_y}).
The factor $R_p^2 I_x + R_s^2 I_y$ in Eq.~(\ref{eq:sgh}) is the intensity of the reflected beam and returns in the expressions of all shifts.
The spatial GH shift is produced by the phase gradients, whereas $R_p$ and $R_s$ can be considered to be small corrections.
Indeed, if we set either $I_x$ or $I_y$ equal to zero in Eq.~(\ref{eq:sgh}), we obtain:
%
\begin{align} 
X_{\mathrm{sGH},x/y} = \dfrac{\lambda}{2 \uppi} \dfrac{\partial \phi_{p/s}}{\partial \theta},
\label{eq:sghx} 	 
\end{align}
which shows that the spatial GH shift consists of two components: $X_{\mathrm{sGH},x}$ for the light polarized in the $x$-direction and $X_{\mathrm{sGH},y}$ for the light polarized in the $y$-direction.
The total spatial GH shift as computed from Eq.~(\ref{eq:sgh}) can then be understood as the intensity-weighted average of these two shifts.

\begin{figure}[!t]
\centering
\includegraphics[width=\hsize]{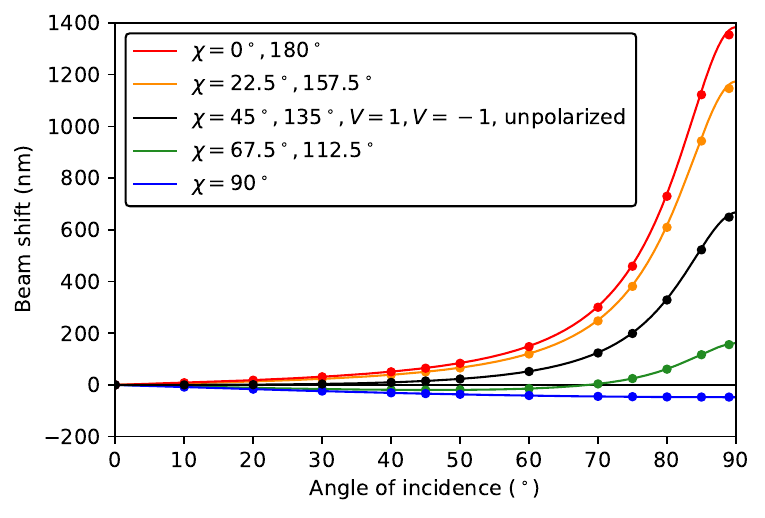} 
\caption{Spatial GH shift as a function of the angle of incidence for reflection off gold at a wavelength of 820~nm as obtained from the closed-form expression of Eq.~(\ref{eq:sgh}) (curves) and polarization ray tracing (data points). The shift is shown for an incident beam of light that is completely unpolarized, 100\% linearly polarized with various angles of linear polarization $\chi$, and 100\% right-handed ($V=1$) or left-handed ($V=-1$) circularly polarized.} 
\label{fig:graph_sgh} 
\end{figure}

Figure~\ref{fig:graph_sgh} shows the spatial GH shift as a function of the angle of incidence for different incident polarization states as computed from Eq.~(\ref{eq:sgh}).
The figure also shows the shifts in the focal plane (data points) as obtained from the numerical computations using the polarization ray tracing as outlined in Sect.~\ref{sec:numerical}.
The close agreement between the analytical and numerical results shows that the spatial GH shift is reproduced very closely by the polarization ray tracing and that Eq.~(\ref{eq:sgh}) is not only valid for Gaussian beams, but is also accurate for beams with a uniform intensity profile. 
Small deviations between the analytical and numerical results are only visible for very large angles of incidence ($\theta \gtrsim 80\degr$).
These deviations are higher-order effects due to the beam intensity profile deviating from a Gaussian profile. 
Indeed, when performing the polarization ray tracing for a Gaussian beam, the data points agree exactly with the analytical curves for all angles of incidence.
%

Figure~\ref{fig:graph_sgh} shows that, although the size of the spatial GH shift is generally less than the wavelength, the shift can be larger than a wavelength for large angles of incidence and certain incident polarization states.
At normal incidence, the shift is always zero.
The spatial GH shift is largest for light polarized in the $x$-direction (i.e.,~for $\chi = 0\degr$ and $\chi = 180\degr$, or $Q = 1$) and increases with increasing angle of incidence.
Because the shift for light polarized in the $x$-direction is directly proportional to $\partial\phi_p / \partial\theta$ (see Eq.~(\ref{eq:sghx})), this behavior can be understood from the increasing gradient seen in Fig.~\ref{fig:fresnel} (right).
For incident light polarized in the $y$-direction (i.e.,~for $\chi = 90\degr$ or $Q = -1$), the shift is much smaller and in the opposite direction, which also agrees with $\partial\phi_s / \partial\theta$ being smaller than and opposite to $\partial\phi_p / \partial\theta$ in Fig.~\ref{fig:fresnel} (right). 
In case of light with $Q = 0$ (e.g.,~for unpolarized light or 100\% polarized light with $\chi = 45\degr$, $\chi = 135\degr$, $V = 1$, or $V = -1$), the intensities of the light polarized in the $x$- and $y$-directions are equal and the resulting shift is the intensity-weighted average of the shifts of the $x$- and $y$-polarizations.
Finally, for light with $0 < \lvert Q \rvert < 1$ (e.g.,~for 100\% polarized light with $\chi = 22.5\degr$, $\chi = 67.5\degr$, $\chi = 112.5\degr$, or $\chi = 157.5\degr$, and also partially polarized light), the resulting shift is in between the three aforementioned shifts.

As can be seen from Fig.~\ref{fig:jones_pupil} (top), which shows the Jones pupil expressed in the $xyz$-basis, the spatial GH shift produces gradients in the phase of all Jones-pupil elements (blue borders).
These phase gradients represent wavefront tilts in the exit pupil and as such result in shifts of the centroid of the PSF in the focal plane.
This confirms the claim by \citet{schmid2018sphere} that the spatial GH shift is the shift that arises from the phase gradient in the $x$-direction in the Jones pupil as described by \citet{breckinridge2015polarization}.
However, we note that Fig.~27 of \citet{schmid2018sphere} suggests that the spatial GH shift is caused by both a shift on the mirror and a directional change of the beam due to a wavefront tilt induced upon reflection.
This depiction is inaccurate: The spatial GH shift is a shift of the entire beam that occurs on the mirror surface, which, in the Fraunhofer approximation, can be described as a wavefront tilt in the exit pupil.

From the Jones pupil, it may seem that the spatial GH shift depends on the f-number, but this is not the case.
Although a two times smaller f-number gives a two times larger phase gradient in the pupil plane, the focal distance is also two times smaller, resulting in the same shift in the focal plane.
Similarly, for a diverging beam (i.e.,~a beam with a negative f-number) the phase gradients have the opposite sign but then the focal plane is virtual and located in front of the mirror (i.e.,~the focal distance is negative), again yielding the same shift.
A more mathematical approach to showing the independence of the shift from the f-number is presented in \citet{schmid2018sphere}. 
We note that the size of the shift (which scales with $\lambda$, see Eq.~(\ref{eq:sgh})) relative to the size of the PSF (which scales with $\lambda \lvert F \rvert$, with $F$ the f-number) does depend on the f-number and is proportional to $1 / \lvert F \rvert$.
This means that a more strongly converging or diverging beam results in a larger shift relative to the PSF.

Finally, we show that the spatial GH shift is visible in the PSM as well (see Fig.~\ref{fig:psm}). 
As described in Sect.~\ref{sec:numerical}, the focal-plane shifts are determined from the intensity image constructed by right-multiplying the PSM by the incident Stokes vector.
In other words, the shifts are determined from the image constructed as a linear combination of the PSM elements in the top row, weighted with the incident Stokes parameters.
Whereas the ${(I\hspace{-2pt}\rightarrow\hspace{-2pt}I)}$-, ${(U\hspace{-2pt}\rightarrow\hspace{-2pt}I)}$-, and ${(V\hspace{-2pt}\rightarrow\hspace{-2pt}I)}$-elements have their centroids shifted in the $x$-direction by the same small amount, the ${(Q\hspace{-2pt}\rightarrow\hspace{-2pt}I)}$-element exhibits a much larger shift in this direction. 
For incident unpolarized light ($Q = U = V = 0$), the shift we find is that of the ${(I\hspace{-2pt}\rightarrow\hspace{-2pt}I)}$-element.
On the other hand, for incident light with $Q$ nonzero, a scaled version of the ${(Q\hspace{-2pt}\rightarrow\hspace{-2pt}I)}$-element, which shows a relatively large shift, is added to or subtracted from the ${(I\hspace{-2pt}\rightarrow\hspace{-2pt}I)}$-element.
This results in a larger, smaller, or opposite shift compared to that of the ${(I\hspace{-2pt}\rightarrow\hspace{-2pt}I)}$-element, in agreement with the curves in Fig.~\ref{fig:graph_sgh}. 
Finally, for incident light with nonzero $U$ and/or $V$, scaled versions of the ${(U\hspace{-2pt}\rightarrow\hspace{-2pt}I)}$- and ${(V\hspace{-2pt}\rightarrow\hspace{-2pt}I)}$-elements are added to or subtracted from the ${(I\hspace{-2pt}\rightarrow\hspace{-2pt}I)}$-element.
However, in this case the resulting shift is the same as that for incident unpolarized light because the centroid shifts of the ${(U\hspace{-2pt}\rightarrow\hspace{-2pt}I)}$- and ${(V\hspace{-2pt}\rightarrow\hspace{-2pt}I)}$-elements are equal to that of the ${(I\hspace{-2pt}\rightarrow\hspace{-2pt}I)}$-element.


\subsection{Angular Goos-H\"anchen shift}
\label{sec:agh}

The angular GH shift, $\varTheta_\mathrm{aGH}$, is an angular deviation of the beam of light upon reflection and, similar to the spatial GH shift, occurs in the plane of incidence \citep[e.g.,][]{aiello2008role, aiello2009duality, gotte2012generalized, bliokh2013goos}.
The definition of the angular GH shift is shown in Fig.~\ref{fig:beam_shifts} (top).
Similar to the spatial GH shift, the angular GH shift can be understood from a 2D picture of the beam of light.
Each ray in the incident beam hits the mirror at a different angle of incidence and therefore experiences a different reflection coefficient.
Over the range of angles of incidence this results in a gradient in the amplitude across the reflected beam (see Fig.~\ref{fig:fresnel}, left), which translates into a shift of the centroid in intensity.
Contrary to the spatial GH shift, the size of the angular shift depends on the divergence angle, and thus the f-number, of the incident beam.
This is because a more strongly converging or diverging beam covers a larger range of angles of incidence and therefore yields a larger gradient. 
The angular GH shift is truly a deflection of the beam centroid as described by an angle, which is the same whether the reflection occurs in the focus or the converging or diverging part of the beam (see Fig.~\ref{fig:beam_shifts}, top).
The resulting physical displacement of the beam centroid vanishes in the focus and increases with distance from the focus. 
That the physical displacement of the beam centroid is zero in the focus can easily be understood in the Fraunhofer approximation: The amplitude gradient in the exit pupil will lead to a point-symmetric change in the PSF, which cannot change the centroid of the intensity distribution.

The angular GH shift can be computed as follows:
\begin{align} 
\varTheta_\mathrm{aGH} = \dfrac{-\alpha^2}{2} \dfrac{R_p \dfrac{\partial R_p}{\partial \theta} I_x + R_s \dfrac{\partial R_s}{\partial \theta} I_y}{R_p^2 I_x + R_s^2 I_y},
\label{eq:agh} 	 
\end{align}  
where, similar to the spatial GH shift, $I_x$ and $I_y$ are functions of Stokes $Q$ (see Eqs.~(\ref{eq:i_x}) and (\ref{eq:i_y})), and $R_p$, $R_s$, and the amplitude gradients $\partial R_p / \partial \theta$ and $\partial R_s / \partial \theta$ (see Fig.~\ref{fig:fresnel}, left) are evaluated at the central angle of incidence.
The divergence angle of the beam, $\alpha$, is given by:
\begin{align} 
\alpha = \arctan\left(\dfrac{1}{2 \lvert F \rvert}\right),
\label{eq:divergence_angle}
\end{align}  
with $F$ the f-number of the beam. 
Contrary to the spatial GH shift, the angular GH shift only depends on the amplitude of the reflection coefficients, and not on the phase.
The angular GH shift is produced by the amplitude gradients, whereas $R_p$ and $R_s$ only have a small effect.
The structure of Eq.~(\ref{eq:agh}) is quite similar to that of Eq.~(\ref{eq:sgh}), which describes the spatial GH shift.
Indeed, when setting $I_x = 0$ or $I_y = 0$ in Eq.~(\ref{eq:agh}), we see that the angular GH shift also consists of two components for the light polarized in the $x$- and $y$-directions:
\begin{align} 
\varTheta_{\mathrm{aGH},x/y} = \dfrac{-\alpha^2}{2} \dfrac{1}{R_{p/s}} \dfrac{\partial R_{p/s}}{\partial \theta}.
\label{eq:aghx} 	 
\end{align}  
Equation~(\ref{eq:agh}) therefore constitutes the intensity-weighted average of these two shifts. 
Finally, the physical displacement of the beam centroid at a distance $z_\mathrm{f}$ from the focus of the beam is given by:
\begin{align} 
X_\mathrm{aGH} = z_\mathrm{f}\,\varTheta_\mathrm{aGH},
\label{eq:aghd} 	 
\end{align}  
where $z_\mathrm{f} > 0$ in the diverging part of the beam and $z_\mathrm{f} < 0$ in the converging part.
We can compute the physical displacement of the centroid of the intensity in the pupil plane by inserting $z_\mathrm{f} = -f$ in Eq.~(\ref{eq:aghd}), where $f$ is the focal distance ($f > 0$ in a converging beam and $f < 0$ in a diverging beam).

\begin{figure}[!t]
\centering
\includegraphics[width=\hsize]{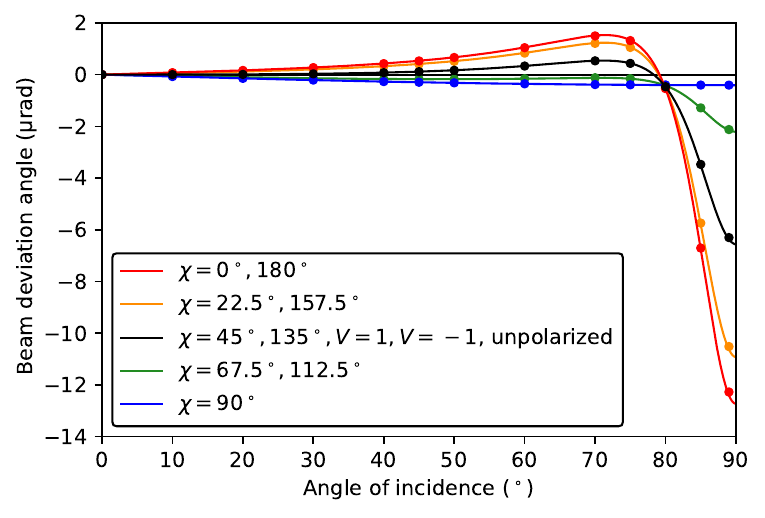}
\caption{Angular GH shift as a function of the angle of incidence at a wavelength of 820~nm for a beam of light with an f-number of 61.3 that reflects off gold as obtained from the closed-form expression of Eq.~(\ref{eq:agh}) (curves) and polarization ray tracing (data points). The shift is shown for an incident beam that is completely unpolarized, 100\% linearly polarized with various angles of linear polarization $\chi$, and 100\% right-handed ($V=1$) or left-handed ($V=-1$) circularly polarized.}
\label{fig:graph_agh} 
\end{figure}

Figure~\ref{fig:graph_agh} shows the angular GH shift as a function of the angle of incidence for different incident polarization states as computed from Eq.~(\ref{eq:agh}).
The figure also shows the shifts as obtained from the exit pupil (data points) using the polarization ray tracing as explained in Sect.~\ref{sec:numerical}.
We have computed these numerical shifts by dividing the physical displacements of the centroid in the pupil plane by the negative value of the focal distance (see Eq.~(\ref{eq:aghd})).
Contrary to the analytically computed shifts, we have computed the numerical shifts only for 100\% polarized light (i.e.,~not for unpolarized light) because the Jones calculus used cannot describe unpolarized or partially polarized light.
Similar to the spatial GH shift, the analytical and numerical results in Fig.~\ref{fig:graph_agh} agree closely and small deviations are only visible for very large angles of incidence.
These deviations are due to the angular GH shift depending on the precise beam intensity profile and vanish when performing the polarization ray tracing for a Gaussian beam.
%

Figure~\ref{fig:graph_agh} indicates that the angular GH shift is on the order of microradians for the particular configuration studied.
For normal incidence, the shift is zero.
The largest shifts are found for light polarized in the $x$-direction (i.e.,~for $\chi = 0\degr$ and $\chi = 180\degr$, or $Q = 1$), whereas the shifts of the light polarized in the $y$-direction (i.e.,~for $\chi = 90\degr$ or $Q = -1$) are much smaller.
The curves can be understood from the amplitude gradients governing the angular GH shift as shown in Fig.~\ref{fig:fresnel} (left): Whereas $\partial R_s / \partial \theta$ increases monotonically with increasing angle of incidence, $\partial R_p / \partial \theta$ is initially negative, reaches a value of zero, and then attains large positive values.
The curves in Fig.~\ref{fig:graph_agh} follow a similar pattern as those of the spatial GH shift (see Fig.~\ref{fig:graph_sgh}), with the shifts for incident light that is not 100\% $x$- or $y$-polarized being an intensity-weighted average of the shifts of the $x$- and $y$-polarizations. 

As shown in the $R_{xx}$- and $R_{yy}$-elements of Fig.~\ref{fig:jones_pupil} (top; red borders), the amplitude gradients associated with the angular GH shift are visible in the Jones pupil expressed in the $xyz$-basis. 
In the antidiagonal elements $R_{xy}$ and $R_{yx}$ these amplitude gradients also exist, but they are overshadowed by the left-right symmetric structure visible in those elements.
For a diverging rather than converging beam, the amplitude gradients have opposite signs (see also Fig.~\ref{fig:beam_shifts}, top). 
Because a diverging beam implies a negative focal distance, that is, the focal plane is virtual and located in front of the mirror, the signs of the angular shifts themselves do not change (see Eq.~(\ref{eq:aghd})).
Finally, the angular GH shift is not visible in the PSM (Fig.~\ref{fig:psm}) because it is zero in the focus.

\vspace{0.5cm}


\subsection{Spatial Imbert-Federov shift}
\label{sec:sif}

The spatial IF shift, $Y_\mathrm{sIF}$, is a displacement of the entire beam of light upon reflection and occurs in the direction perpendicular to the plane of incidence \citep[e.g.,][]{federov1955k, imbert1972calculation, bliokh2006conservation, aiello2008role, hermosa2011spin, gotte2012generalized, bliokh2013goos, bliokh2015spin}. 
A schematic with the definition of the spatial IF shift is shown in Fig.~\ref{fig:beam_shifts} (bottom).
Similar to the spatial GH shift, the spatial IF shift is independent of the f-number of the beam and the position within the beam where the reflection occurs.
To understand the spatial IF shift from a plane-wave decomposition, it is necessary to consider the full 3D picture \citep[e.g.,][]{aiello2008role, bliokh2013goos}.
Each plane wave in the incident beam has a different (3D) propagation direction.
Therefore, not only the angles of incidence (and thus the reflection coefficients) are different among the waves, but also the orientations of the local planes of incidence.
These rotations of the planes of incidence induce different geometric (Berry) phases among the circularly polarized components of the waves.
This results in a gradient of the geometric phases in the direction perpendicular to the plane of incidence, with the gradient having opposite sign for the right-handed and left-handed circular polarizations.
Accounting for the reflection coefficients of each wave as well as the geometric phases within the reflected beam, the reflected beam is found to be shifted in the direction perpendicular to the plane of incidence when integrating over all waves.

The spatial IF shift is more easily understood in terms of conservation of total angular momentum \citep[e.g.,][]{bliokh2006conservation, bliokh2013goos, bliokh2015transverse, bliokh2015spin}.
Disregarding vortex beams, the total angular momentum of a beam of light consists of the spin angular momentum (SAM) and the external orbital angular momentum.
In the quantum-mechanical description of light, photons carry one of two spin states that correspond to right-handed and left-handed circular polarization.
The SAM of a beam of light is a vector quantity pointing in the direction of propagation that is proportional to the difference between the number of right-handed and left-handed photons, that is, it is proportional to Stokes $V$.
The external orbital angular momentum is given by the cross product of the radius vector of the beam centroid with respect to some origin and the linear momentum of the beam, with the latter pointing in the direction of propagation.
Upon reflection, the total angular momentum in the direction normal to the surface of the mirror is conserved.
As a result, any change in the SAM of the beam, that is, in the circular polarization, must be compensated for by a shift of the beam in the direction perpendicular to the plane of incidence.
This shift is the spatial IF shift, which is therefore considered to be a spin-orbit interaction of light. 

The spatial IF shift can be calculated as follows:
\begin{align} 
Y_\mathrm{sIF} = \dfrac{-\lambda}{2 \uppi} \dfrac{\cot\theta}{R_p^2 I_x + R_s^2 I_y} \left[V \left(\dfrac{R_p^2 + R_s^2}{2} \right) + R_p R_s \left(V \cos\varDelta + U \sin\varDelta \right) \right],
\label{eq:sif} 	 
\end{align}  
where $R_p$, $R_s$, and the retardance $\varDelta$ (see Eq.~(\ref{eq:retardance}) and Fig.~\ref{fig:fresnel}) are evaluated at the central angle of incidence $\theta$, and $\cot\theta$ is the transverse gradient of the induced geometric phase. 
Although the spatial IF shift has a weak dependence on Stokes $Q$ through $I_x$ and $I_y$ (see Eqs.~(\ref{eq:i_x}) and (\ref{eq:i_y})), the shift depends primarily on the incident Stokes $U$ and $V$.
So, whereas the GH shift consists of two separate shifts for light polarized in the $x$- and $y$-directions, the spatial IF shift comprises separate and opposite shifts for the diagonally and antidiagonally polarized components (because $U = I_d - I_a$, see Eq.~(\ref{eq:stokes_u})) as well as for the right-handed and left-handed circularly polarized components (because $V = I_r - I_l$, see Eq.~(\ref{eq:stokes_v})).
For metallic reflections, the spatial IF shift results primarily from the retardance, whereas $R_p$ and $R_s$ can be considered to be small corrections.
Indeed, we can simplify Eq.~(\ref{eq:sif}) by assuming that the incident beam is totally reflected.
Setting $R_p = R_s = 1$ and inserting $I_x + I_y = 1$ (see Eq.~(\ref{eq:stokes_i})), we obtain: 
\begin{align} 
Y_\mathrm{sIF} = \dfrac{-\lambda}{2 \uppi} \cot\theta ~\Bigl[ V \left(1 + \cos\varDelta \right) + U \sin\varDelta \Bigr].
\label{eq:sifx} 	 
\end{align}  
In this equation, the factor $[V(1 + \cos\varDelta) + U \sin\varDelta]$ is proportional to the change of the SAM upon reflection, with $V(1)$ proportional to the incident SAM and $-(V\cos\varDelta + U \sin\varDelta)$, which gives Stokes $V$ after reflection, proportional to the SAM of the reflected beam.
The spatial IF shift thus depends on the crosstalk from $U$ to $V$ ($U \sin\varDelta$) and the crosstalk from $V$ to $U$ or even the crosstalk creating a change of handedness of the circular polarization ($V\cos\varDelta$).

\begin{figure}[!t]
\centering
\includegraphics[width=\hsize]{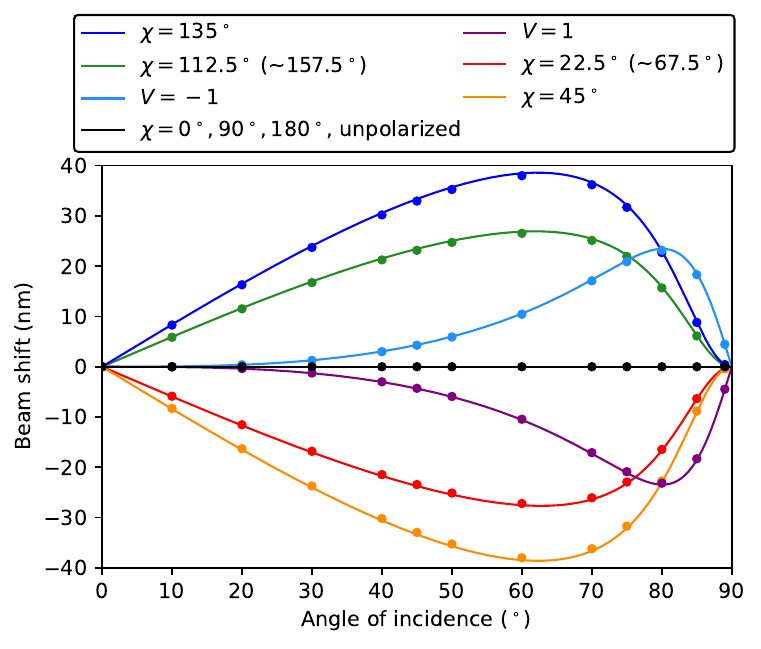} 
\caption{Spatial IF shift as a function of the angle of incidence for reflection off gold at a wavelength of 820~nm as obtained from the closed-form expression of Eq.~(\ref{eq:sif}) (curves) and polarization ray tracing (data points). The shift is shown for an incident beam of light that is completely unpolarized, 100\% linearly polarized with various angles of linear polarization $\chi$, and 100\% right-handed ($V=1$) or left-handed ($V=-1$) circularly polarized. The shifts for $\chi = 67.5\degr$ and $\chi = 157.5\degr$ are not shown, but are very close to the shifts for $\chi = 22.5\degr$ and $\chi = 112.5\degr$, respectively. The colors indicate different polarization states than in Figs.~\ref{fig:graph_sgh} and \ref{fig:graph_agh}.} 
\label{fig:graph_sif} 
\end{figure}

Figure~\ref{fig:graph_sif} shows the spatial IF shift as a function of the angle of incidence for different incident polarization states as computed from Eq.~(\ref{eq:sif}).
Also shown are the shifts in the focal plane (data points) as numerically determined using polarization ray tracing (see Sect.~\ref{sec:beam_shifts}), which agree closely with the analytical computations.
The small deviations among the results vanish when performing the polarization ray tracing with a Gaussian beam.
%

Figure~\ref{fig:graph_sif} illustrates that the spatial IF shift is (somewhat) smaller than the spatial GH shift and is always smaller than the wavelength. 
At normal incidence, where $\varDelta = 180\degr$ (see Fig.~\ref{fig:fresnel}), the spatial IF shift is zero.
For nonzero angles of incidence, where $\varDelta \neq 180\degr$, changes in the SAM occur for incident $U$- or $V$-polarized light, thus leading to spatial IF shifts.
The spatial IF shifts are in opposite directions for opposite signs of $U$ (e.g.,~for $\chi = 45\degr$ and $\chi = 135\degr$) and $V$ (for right-handed and left-handed circular polarization).
The shifts initially become larger with increasing angle of incidence (because $\varDelta$ monotonically decreases), but then become smaller again for (very) large angles of incidence as $\cot\theta \to 0$ when $\theta \to 90\degr$, resulting in no shift at $\theta = 90\degr$.
The spatial IF shift for $U$ ($\chi = 45\degr$ and $\chi = 135\degr$) reaches larger values than that of $V$ with the maximum of $U$ occurring at a smaller angle of incidence than the maximum of $V$.
The maxima of the curves are lower for partially polarized light or light with both $Q$ and $U$ nonzero (e.g.,~$\chi = 22.5\degr$, $\chi = 67.5\degr$, $\chi = 112.5\degr$, or $\chi=157.5\degr$).
Although the light with $\chi = 22.5\degr$ and $\chi = 67.5\degr$ (and similar for $\chi = 112.5\degr$ and $\chi=157.5\degr$) have the same value for $U$, small differences in the size of the shifts occur due to the dependence on $Q$ via $I_x$ and $I_y$.
The curves of incident light with both $U$ and $V$ nonzero (not shown in Fig.~\ref{fig:graph_sif}) are combinations of the curves for the individual Stokes parameters.
Finally, for unpolarized light or light polarized in the $x$- or $y$-direction (i.e.,~$Q$-polarized light), the spatial IF shift is always zero because the incident beam overall carries no SAM and no SAM can be created upon reflection.

Similar to the spatial GH shift, the spatial IF shift is expected to create gradients in phase in the Jones pupil.
However, in the Jones pupil expressed in the $xyz$-basis (see Fig.~\ref{fig:jones_pupil}, top), phase gradients in the $y$-direction are not visible.
This is because the spatial IF shift primarily depends on Stokes $U$ and $V$ (see Eq.~(\ref{eq:sifx})), and therefore results from the complex linear combination of all four Jones-pupil elements in this basis. 
Nevertheless, a hint of a gradient in the $y$-direction is visible in the $R_{xy}$-, $R_{yx}$-, $\phi_{xy}$-, and $\phi_{yx}$-elements when considering that a phase difference of $\uppi$ between the left and right sides of the pupil implies that the reflection coefficients on either side have opposite signs.
Actual phase gradients in the $y$-direction naturally appear in the Jones pupils expressed in the bases of Stokes $U$ and $V$, that is, in the Jones pupils expressed in the $daz$- and $rlz$-bases (see Fig.~\ref{fig:jones_pupil}, center and bottom).
The gradients are visible in the \mbox{$\phi_{da}$-,} \mbox{$\phi_{ad}$-,} $\phi_{rr}$-, and $\phi_{ll}$-elements (green borders).
The Jones pupils also show the phase gradient in the $x$-direction produced by the spatial GH shift (blue borders), with the $\phi_{da}$- and $\phi_{ad}$-elements exhibiting a combination of gradients in the $x$- and $y$-directions.
In Fig.~\ref{fig:jones_pupil} (center and bottom), the amplitude gradient in the $x$-direction due to the angular GH is visible as well (red borders).
Lastly, we note that although the spatial IF shift does not depend on the f-number, its size relative to the PSF scales as $1 / \lvert F \rvert$, analogous to the spatial GH shift (see Sect.~\ref{sec:sgh}).


Finally, we show how the spatial IF shift is visible in the PSM (see Fig.~\ref{fig:psm}).
As explained in Sect.~\ref{sec:sgh}, the focal-plane shifts are determined from the image created as a linear combination of the PSM elements in the top row, weighted with the incident Stokes parameters.
Because the ${(I\hspace{-2pt}\rightarrow\hspace{-2pt}I)}$- and ${(Q\hspace{-2pt}\rightarrow\hspace{-2pt}I)}$-elements are symmetric with respect to the $x$-axis (i.e.,~they are left-right symmetric in Fig.~\ref{fig:psm}), no shift results for unpolarized light or light that is polarized in the $x$- or $y$-direction.
The ${(U\hspace{-2pt}\rightarrow\hspace{-2pt}I)}$- and ${(V\hspace{-2pt}\rightarrow\hspace{-2pt}I)}$-elements on the other hand are asymmetric, with positive and negative signals on opposite sides of the $x$-axis.
For incident light with nonzero $U$ and/or $V$, scaled versions of these elements are added to or subtracted from the ${(I\hspace{-2pt}\rightarrow\hspace{-2pt}I)}$-element, producing a PSF with the centroid shifted in the $y$-direction. 
We note that the relative intensity of the ${(U\hspace{-2pt}\rightarrow\hspace{-2pt}I)}$-element is larger than that of the ${(V\hspace{-2pt}\rightarrow\hspace{-2pt}I)}$-element, in agreement with the spatial IF shift being larger for $U$ than for $V$ at an angle of incidence of $45\degr$ (see Fig.~\ref{fig:graph_sif}).

\vspace{1cm}


\subsection{Angular Imbert-Federov shift}
\label{sec:aif}

The angular IF shift, $\varTheta_\mathrm{aIF}$, is an angular deviation of the beam of light upon reflection directed away from the plane of incidence \citep[e.g.,][]{bliokh2007polarization, aiello2008role, hermosa2011spin, gotte2012generalized, bliokh2013goos}.
The definition of the angular IF shift is shown in Fig.~\ref{fig:beam_shifts} (bottom).
The angular IF shift is related to the conservation of linear momentum in the direction perpendicular to the plane of incidence, and, similar to the spatial IF shift, results from the differences in induced geometric phase across the beam.
Similar to the angular GH shift, the size of the angular IF shift depends on the f-number of the incident beam and is the same whether the beam is reflected in the focus or in the converging or diverging parts of the beam.
The physical displacement of the centroid of the beam is zero in the focus and increases with distance from the focus. 

The angular IF shift can be calculated as follows:
\begin{align} 
\varTheta_\mathrm{aIF} = \dfrac{\alpha^2}{4} \dfrac{\cot\theta}{R_p^2 I_x + R_s^2 I_y} U \left(R_p^2 - R_s^2\right),
\label{eq:aif} 	 
\end{align}  
where $R_p$ and $R_s$ are computed at the central angle of incidence, and the divergence angle $\alpha$ is given by Eq.~(\ref{eq:divergence_angle}). 
Similar to the angular GH shift, the angular IF shift does not depend on the phases of the reflection coefficients, but only on the amplitudes.
Although the angular IF shift has small $Q$-dependent corrections through $I_x$ and $I_y$ (see Eqs.~(\ref{eq:i_x}) and (\ref{eq:i_y})), the shift depends primarily on the incident Stokes $U$.
The angular IF shift consists of separate and opposite shifts for the
diagonally and antidiagonally polarized components (because $U = I_d - I_a$, see Eq.~(\ref{eq:stokes_u})) and results primarily from the diattenuation. 
Indeed, if $Q = 0$, that is, $I_x = I_y = {}^1{\mskip -3mu/\mskip -1mu}_2$, Eq.~(\ref{eq:aif}) reduces to:
\begin{align} 
\varTheta_\mathrm{aIF} = \dfrac{-\alpha^2}{2} U \epsilon \cot\theta,
\label{eq:aifx} 	 
\end{align}  
with $\epsilon$ the diattenuation from Eq.~(\ref{eq:diattenuation}).
Finally, the physical displacement of the centroid of the beam is given by:
\begin{align} 
Y_\mathrm{aIF} = z_\mathrm{f}\,\varTheta_\mathrm{aIF},
\label{eq:aifd} 	 
\end{align}  
with $z_\mathrm{f}$ the distance from the focus, similar to Eq.~(\ref{eq:aghd}).

\begin{figure}[!t]
\centering
\includegraphics[width=\hsize]{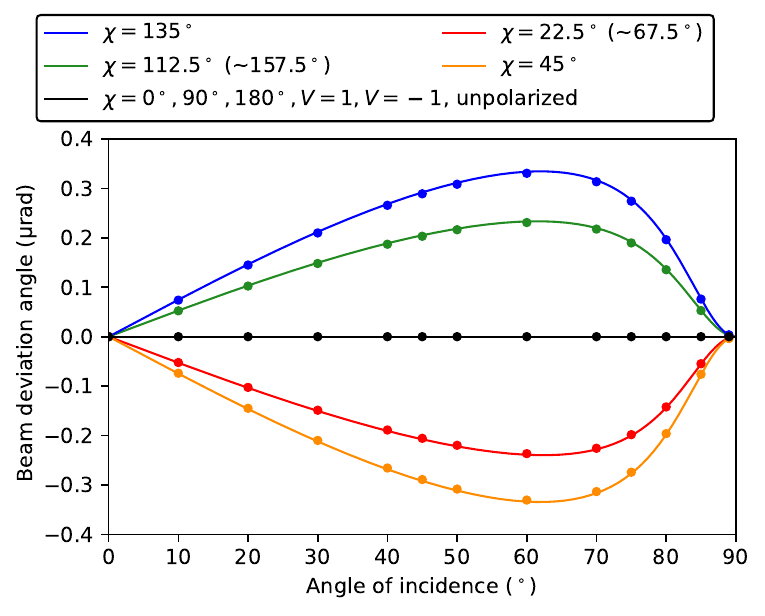} 
\caption{Angular IF shift as a function of the angle of incidence at a wavelength of 820~nm for a beam of light with an f-number of 61.3 that reflects off gold as obtained from the closed-form expression of Eq.~(\ref{eq:aif}) (curves) and polarization ray tracing (data points). The shift is shown for an incident beam that is completely unpolarized, 100\% linearly polarized with various angles of linear polarization $\chi$, and 100\% right-handed ($V=1$) or left-handed ($V=-1$) circularly polarized. The shifts for $\chi = 67.5\degr$ and $\chi = 157.5\degr$ are not shown, but are very close to the shifts for $\chi = 22.5\degr$ and $\chi = 112.5\degr$, respectively. Except for the circular polarization, the colors used indicate the same polarization states as in Fig.~\ref{fig:graph_sif}.} 
\label{fig:graph_aif} 
\end{figure}

Figure~\ref{fig:graph_aif} shows the angular IF shift as a function of the angle of incidence for different incident polarization states as computed from Eq.~(\ref{eq:aif}).
The shifts as obtained from the exit pupil (data points) using polarization ray tracing (see Sect.~\ref{sec:numerical}) are also shown.
These numerical shifts are computed using Eq.~(\ref{eq:aifd}) and are only calculated for 100\% polarized light, similarly to the angular GH shifts (see Sect.~\ref{sec:agh}).
The analytical and numerical results agree closely, with the small deviations vanishing when performing the polarization ray tracing for a Gaussian beam.
%

Figure~\ref{fig:graph_aif} shows that the angular IF shift is on the order of less than a microradian for the particular configuration considered.
For incident light with $U$ nonzero, angular IF shifts occur that are in the opposite direction for opposite signs of $U$.
The shifts are zero for angles of incidence of $0\degr$ and $90\degr$.
The shape of the curves is related to the diattenuation (roughly the difference between $R_s$ and $R_p$ in Fig.~\ref{fig:fresnel}), which initially increases with increasing angle of incidence, reaches a maximum, and then decreases again to zero at $\theta = 90\degr$.
For incident light with $U = 0$ (i.e.,~$\chi = 0\degr$, $\chi = 90\degr$, $\chi = 180\degr$, $V = 1$, $V = -1$, or unpolarized light), the shift is zero for any angle of incidence.

Finally, the amplitude gradients in the $y$-direction associated with the angular IF shift are visible in the $R_{da}$- and $R_{ad}$-elements of the Jones pupil expressed in the $daz$-basis (see Fig.~\ref{fig:jones_pupil}, center).
The gradients of these elements are a combination of gradients in the $y$-direction and the $x$-direction, with the latter due to the angular GH shift (red borders).
Because the angular IF shift is zero in the focus, it is not visible in the PSM. 

%
%

\section{Discussion}
\label{sec:discussion}

In Sect.~\ref{sec:beam_shifts} we explained the origin and characteristics of the spatial and angular GH and IF shifts and investigated their size and direction as a function of the angle of incidence and incident polarization state.
We also showed that all four beam shifts are fully reproduced by polarization ray tracing as described in Sect.~\ref{sec:numerical} and that the exact beam intensity profile (i.e.,~whether it is Gaussian or uniform) has a negligible effect. 
Of the four beam shifts, only the spatial GH and IF shifts are relevant for high-contrast imagers and telescopes because these shifts are visible in the focal plane; the angular GH and IF shifts are not important because, besides a small point-symmetric deformation of the PSF for angles of incidence close to grazing incidence (which do not occur in high-contrast imagers), they have no effect in the focus.
We thus find that the polarization structure in the PSF that limits the performance of coronagraphs and the speckle suppression of polarimetric imagers is created by the spatial GH and IF shifts.
We note that the effect of these shifts on high-resolution spectroscopy and astrometry of planets should generally be small.
The fiber-positioning system of a high-resolution spectrograph maximizes the amount of planet light that enters the fiber, thereby automatically correcting for the beam shifts.
And because the beam shifts are similar for astronomical objects at different locations on the science detector, relative astrometry is almost not affected.

In Sect.~\ref{sec:polarization_structure} we investigate the polarization structure in the PSF created by the spatial GH and IF shifts.
Subsequently, in Sect.~\ref{sec:beam_shifts_polarimetry}, we examine the effect of the spatial GH and IF shifts on polarimetric measurements.
In Sect.~\ref{sec:wavelength_material} we then briefly discuss the size of the spatial GH and IF shifts for various mirror materials and wavelengths.
After that, in Sect.~\ref{sec:mitigation}, we use our understanding of the spatial GH and IF shifts to discuss and refine the approaches to mitigate the shifts.   
Finally, we present a table summarizing the properties of the four beam shifts in Sect.~\ref{sec:overview_beam_shifts}.


\subsection{Polarization structure in the PSF due to beam shifts}
\label{sec:polarization_structure}

In this section, we investigate the polarization structure in the PSF created by the spatial GH and IF shifts.
This polarization structure must be taken into account when designing the coronagraphs of high-contrast imagers that aim to detect planets in reflected light \citep{breckinridge2015polarization}.
For our analysis, we consider the reflection off a single flat mirror at an angle of incidence of $45\degr$, using the same configuration as examined in Sects.~\ref{sec:numerical} and \ref{sec:beam_shifts}.

The observed light of the stars around which high-contrast imagers search for planets is unpolarized or has a degree of polarization of only several percent \citep[see e.g.,][]{heiles20009286}.
For this case of (nearly) unpolarized incident light, the Stokes vector after reflection off a flat mirror is given by the elements in the left column of the PSM in Fig.~\ref{fig:psm}, that is, the ${(I\hspace{-2pt}\rightarrow\hspace{-2pt}I)}$-, ${(I\hspace{-2pt}\rightarrow\hspace{-2pt}Q)}$-, ${(I\hspace{-2pt}\rightarrow\hspace{-2pt}U)}$-, and ${(I\hspace{-2pt}\rightarrow\hspace{-2pt}V)}$-elements.
These elements are the same as those in the top row of the PSM, except for the ${(I\hspace{-2pt}\rightarrow\hspace{-2pt}U)}$-element which has opposite sign.
Because the spatial GH and IF shifts follow from these top-row elements (see Sects.~\ref{sec:sgh} and \ref{sec:sif}), the polarization-dependent structures visible in the Stokes vector for reflection of incident unpolarized light must be created by the spatial GH and IF shifts.
In the following, we refer to the \mbox{${(I\hspace{-2pt}\rightarrow\hspace{-2pt}I)}$-,} \mbox{${(I\hspace{-2pt}\rightarrow\hspace{-2pt}Q)}$-,} ${(I\hspace{-2pt}\rightarrow\hspace{-2pt}U)}$-, and ${(I\hspace{-2pt}\rightarrow\hspace{-2pt}V)}$-elements as the intensity image and the $Q$-, $U$-, and $V$-images, respectively.

As outlined in Sect.~\ref{sec:sgh}, the spatial GH shift is described by two opposite shifts of different size for the incident light polarized in the $x$- and $y$-directions, that is, for the incident $I_x$- and $I_y$-components of the light.
Because unpolarized light can be described as the sum of equal amounts of the $I_x$- and $I_y$-components 
(see Eqs.~(\ref{eq:stokes_i}), (\ref{eq:i_x}), and (\ref{eq:i_y})), the intensity image consists of two PSF components that are slightly shifted in opposite directions along the $x$-axis.
As a result, the PSF in intensity is not only shifted (by $15$~nm or 1.8\% of the wavelength for the configuration considered; see Fig.~\ref{fig:graph_sgh}, black curve), but also broadened in the $x$-direction.
The $Q$-image is equal to the difference of the $I_x$- and $I_y$-components (see Eq.~(\ref{eq:stokes_q})).
Due to the diattenuation (see Eq.~(\ref{eq:diattenuation})), the two components are not reflected by an equal amount.
Therefore, an overall negative signal with a minimum of ${\sim}0.9\%$ remains in the image, which constitutes the instrumental polarization.
But because the $I_x$- and $I_y$-components are also shifted in opposite directions, this instrumental-polarization signal itself also has a large shift \citep[see also][]{breckinridge2015polarization}. 

As explained in  Sect.~\ref{sec:sif}, the spatial IF shift is opposite for incident diagonally ($d$) and antidiagonally ($a$) polarized light (i.e.,~for positive and negative 100\% $U$-polarized light) as well as for incident right-handed ($r$) and left-handed ($l$) circularly polarized light (i.e.,~for positive and negative 100\% $V$-polarized light).
Unpolarized light can be described as the sum of equal amounts of these $I_d$- and $I_a$-components as well as the sum of equal amounts of the $I_r$- and $I_l$-components (see Eqs.~(\ref{eq:stokes_i}), (\ref{eq:stokes_u}), and (\ref{eq:stokes_v})).
Therefore, the intensity image consists of PSF components that are slightly shifted by equal amounts in opposite directions parallel to the $y$-axis.
So although the PSF in intensity is not shifted by the spatial IF shift when the incident light is unpolarized (see Fig.~\ref{fig:graph_sif}, black curve), it is broadened in the $y$-direction in addition to the broadening in the $x$-direction (due to the spatial GH shift).
The opposite shifts of the $I_d$- and $I_a$-components and the $I_r$- and $I_l$-components can also be seen in the $U$- and $V$-images, respectively, where they create asymmetric structures with positive and negative signals on opposite sides of the $x$-axis.
For the configuration considered, these structures have values below $0.1\%$ of the intensity (with the $U$-image having larger values than the $V$-image as can be expected from Fig.~\ref{fig:graph_sif}). 
The asymmetric structures are also visible in the $R'_{xy}$-, $R'_{yx}$-, $\phi'_{xy}$-, and $\phi'_{yx}$-elements of the ARM (see Fig.~\ref{fig:arm}). 
\citet{breckinridge2015polarization} refer to these structures in the ARM as ghost PSFs (see Sect.~\ref{sec:introduction}).
Our results therefore show that these ghost PSFs are created by the spatial IF shifts and are elliptically polarized.
Finally, we note that due to the splitting of the orthogonal circular polarization states in the $V$-image, the spatial IF shift is often also referred to as the spin Hall effect of light \citep[e.g.,][]{hermosa2011spin, bliokh2013goos, bliokh2015transverse, bliokh2015spin}.

The PSM in Fig.~\ref{fig:psm} as calculated with polarization ray tracing includes all orders of polarization aberrations.
Still, we find that the polarization structure in the PSF for the case of incident unpolarized light is adequately described by the diattenuation (i.e.,~the instrumental polarization) and the first-order polarization aberrations in the focus, that is, the spatial GH and IF shifts.
We therefore conclude that only for curved mirrors the higher-order polarization aberrations, such as polarization-dependent astigmatism \citep{breckinridge2015polarization}, come into play.
For a discussion on the combined effect of a series of flat mirrors and the polarization aberrations of curved mirrors with normal incidence, we refer to \citet{breckinridge2015polarization}.


\subsection{Effect of beam shifts on polarimetric measurements}
\label{sec:beam_shifts_polarimetry}

In this section, we investigate the effect of the spatial GH and IF shifts on polarimetric measurements with high-contrast imagers.
The physics literature does not describe beam shifts for the case where unpolarized light is incident on a mirror and where the reflected light is subsequently measured by a polarimeter. 
However, we can understand this case from our insights into the beam shifts and our results from the polarization ray tracing.

We shall consider a rotatable linear polarizer placed behind the mirror that we analyzed in Sect.~\ref{sec:polarization_structure}.
In that case, the Stokes vector incident on the polarizer is the same Stokes vector as examined in Sect.~\ref{sec:polarization_structure}: It is equal to the left column of the PSM in Fig.~\ref{fig:psm}.
If we then align the transmission axis of the polarizer with the $x$-, $y$-, $d$-, and $a$-directions, we measure the $I_x$-, $I_y$-, $I_d$-, and $I_a$-components of the beam. 
Also, if we replace the polarizer with a right-handed or left-handed circular polarizer, we measure the $I_r$- and $I_l$-components of the beam.
As a result, these six measurements are sensitive to exactly the same spatial GH and IF shifts of these components as described in Sect.~\ref{sec:polarization_structure}.
Therefore, when we compute the differences of the $x$- and $y$-, $d$- and $a$-, and $r$- and $l$-measurements, we obtain the $Q$-, $U$-, and $V$-images of the Stokes vector after reflection.

Because stars are generally unpolarized, polarimetric measurements strongly suppress the light from the star, thereby making the detection of planets in reflected light easier. 
However, the maximum gain in contrast from polarimetry is limited by the spatial GH and IF shifts and the polarization structure that they create.
Although the instrumental polarization is a larger aberration, this effect is routinely subtracted in the data reduction and/or removed by using a half-wave plate in front of the optical path in current high-contrast imaging polarimeters \citep{witzel2011instrumental, canovas2011data, wiktorowicz2014gemini, millar2016gpi, de2020polarimetric, van2020calibration, van2020polarimetric}.

\begin{figure}[!t]
\centering
\includegraphics[trim=30pt 5pt 10pt 5pt, clip, width=\hsize]{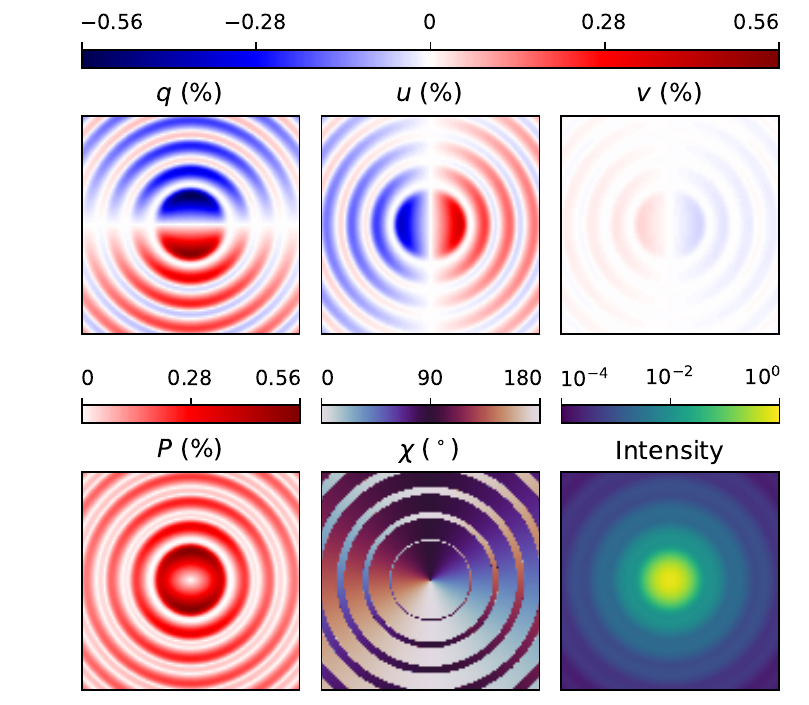}
\caption{Images of the PSF structures visible in normalized Stokes $q$ (without instrumental polarization), $u$, and $v$ (\emph{top}), degree of linear polarization $P$, angle of linear polarization $\chi$, and intensity (\emph{bottom}) at a wavelength of 820~nm for a converging beam of light with an f-number of 61.3 that reflects off gold at an angle of incidence of $45\degr$.
The images are convolved with a top-hat kernel with a diameter equal to the full width at half maximum of the PSF in intensity.
The images show the core of the PSF and the first three complete Airy rings.
The positive $x$- and $y$-directions are upward and to the left, respectively.
} 
\label{fig:normalized_stokes_elements} 
\end{figure}

To quantify the maximum gain in contrast from polarimetry as limited by the spatial GH and IF shifts, we compute the mirror-induced fractional polarization in $Q$, $U$, and $V$ over the PSF.
To this end, we convolve the intensity image and the $Q$-, $U$-, and $V$-images using a top-hat kernel with a diameter equal to the full width at half maximum of the PSF in the intensity image.
This diameter is equal to the diameter of the apertures one would use to extract the fluxes of detected planets and determine the noise level in the images \citep[e.g.,][]{mawet2014fundamental}.
After convolving the images, we compute the instrumental polarization in the $Q$-image by dividing the total flux in the $Q$-image by the total flux in the intensity image.
We then subtract the instrumental polarization from the $Q$-image by multiplying the intensity image by the instrumental polarization and subtracting the resulting image from the $Q$-image.
Subsequently, we compute the images of the normalized Stokes parameters $q = Q/I$, $u = U/I$, and $v = V/I$ by dividing the (instrumental-polarization-subtracted) $Q$-, $U$-, and $V$-images by the intensity image.
The resulting images as well as the images of the intensity and the degree and angle of linear polarization $P$ and $\chi$ (see Eqs.~(\ref{eq:dolp}) and (\ref{eq:aolp})) are shown in Fig.~\ref{fig:normalized_stokes_elements}.
%

Figure~\ref{fig:normalized_stokes_elements} (top) shows that the spatial GH and IF shifts create a significant polarization structure in the PSF. 
In all images the PSF core and the Airy rings show an asymmetric structure with successive positively and negatively polarized regions.
In case of the $u$- and $v$-images, we found that these structures are created by the spatial IF shifts and identified them as the ghost PSFs described by \citet{breckinridge2015polarization} (see Sect.~\ref{sec:polarization_structure}).
However, by subtracting the instrumental polarization, we have revealed an even stronger asymmetric structure or ghost PSF in the $q$-image.
In this case, the structure is produced by the spatial GH shifts and is oriented orthogonally to the structures in the $u$- and $v$-images.

Figure~\ref{fig:normalized_stokes_elements} (top) also shows that the PSF has significant fractional-polarization levels, with the largest values in the $q$-image and the smallest values in the $v$-image.
The relative strengths of the fractional polarization in the $q$-, $u$-, and $v$-images are directly related to the relative sizes of the spatial GH and IF shifts at an angle of incidence of $45\degr$ (see Figs.~\ref{fig:graph_sgh} and \ref{fig:graph_sif}).
Figure~\ref{fig:normalized_stokes_elements} (bottom) indicates that the degree of linear polarization in the PSF reaches a maximum of 0.56\%.
Finally, we see that the angle of linear polarization rotates $180\degr$ when moving in a circle around the center of the PSF and that it differs by $90\degr$ between the inner and outer regions of the Airy rings. 

The polarization structure in the $q$-, $u$-, and $v$-images limit the local gain in contrast achievable with polarimetry.
The degree of (linear) polarization is several tenths of a percent on average; hence the average contrast gain is a factor of ${\sim}350$, which is the gain compared to the contrast in intensity including the effects of seeing.
This is because any speckles due to the seeing are also polarized at approximately this level.
We stress that the exact numerical values presented in Fig.~\ref{fig:normalized_stokes_elements} are only valid for the specific configuration considered.
For example, for a series of mirrors and/or beams with smaller f-numbers, the fractional-polarization levels are much higher and therefore the gain in contrast due to polarimetry is much lower.

Finally, as discussed in Sect.~\ref{sec:introduction}, the polarimetric speckle suppression of the high-contrast imaging polarimeter SPHERE-ZIMPOL is limited by polarization-dependent beam shifts \citep{schmid2018sphere}. 
Indeed, the structures visible in the on-sky polarimetric images of Fig.~26 of \citet{schmid2018sphere} agree very well with the asymmetric structures (ghost PSFs) in the $q$- and $u$-images of Fig.~\ref{fig:normalized_stokes_elements} (top).
Therefore, the polarimetric contrast of SPHERE-ZIMPOL at small angular separations from the star is clearly limited by both the spatial GH and IF shifts.


\subsection{Size of beam shifts for various mirror materials and wavelengths}
\label{sec:wavelength_material}

\begin{figure}[!t]
\centering
\includegraphics[width=\hsize]{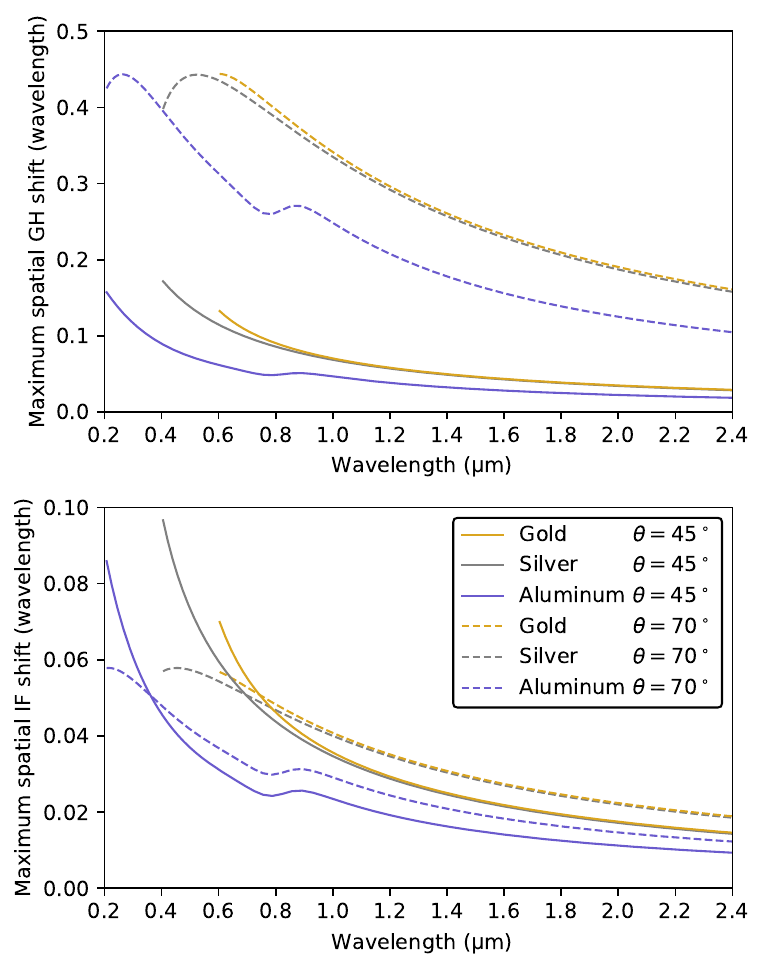} 
\caption{Maximum wavelength-normalized spatial GH (\emph{top}) and IF (\emph{bottom}) shifts as a function of wavelength at an angle of incidence $\theta$ of $45\degr$ and $70\degr$ for reflection off gold, silver, and aluminum. The legend in the bottom panel is valid for both panels. The shifts for gold and silver are only shown for wavelengths longer than 600~nm and 400~nm, respectively, because the reflectivity drops below 90\% at shorter wavelengths.} 
\label{fig:graph_beam_shift_wavelength} 
\end{figure}

So far we have only considered the beam shifts for reflection off gold at a wavelength of 820~nm.
Here we briefly discuss the maximum size of the spatial GH and IF shifts as a function of wavelength from the ultraviolet to the near-infrared for the three most common (bulk) mirror materials used in astronomical telescopes and instruments.
We note, however, that actual mirrors in astronomical telescopes and instruments are likely to consist of a stack of thin films and so the exact sizes of the shifts will be different. 
To compute the shifts, we use the complex refractive indices of gold, silver, and aluminum for the range of wavelengths from \citet{rakic1998optical}.
Figure~\ref{fig:graph_beam_shift_wavelength} shows the spatial GH shift for $x$-polarized light (from Eq.~(\ref{eq:sgh})) and the spatial IF shift for antidiagonally polarized light (from Eq.~(\ref{eq:sif})), both normalized with the wavelength, for angles of incidence $\theta$ equal to $45\degr$ and $70\degr$.
%

Figure~\ref{fig:graph_beam_shift_wavelength} shows that the spatial GH shift is larger than the spatial IF shift for all mirror materials, that the size of the shifts is always less than the wavelength, and that the shifts relative to the wavelength are larger for shorter wavelengths.
Of the three materials, aluminum produces the smallest shifts, whereas gold and silver create larger shifts.
For all materials and wavelengths, the spatial GH shift is smaller for $\theta = 45\degr$ than for $\theta = 70\degr$.
The same is true for the spatial IF shift, except for the shortest wavelengths where the shift for $\theta = 45\degr$ becomes larger than that for $\theta = 70\degr$.
  

\subsection{Mitigation of beam shifts}
\label{sec:mitigation}

\citet{breckinridge2015polarization} provide possible approaches to mitigate polarization aberrations in optical systems, which include using beams of light with large f-numbers, keeping the angles of incidence small, and tuning the coatings of the mirrors.
In this section, we discuss and refine these approaches based on our fundamental understanding of the beam shifts.
\citet{breckinridge2015polarization} also discuss the use of possible optical devices that could compensate polarization aberrations \citep[see also][]{clark2011polarization, sit2017general, dai2019adaptive}, but a discussion of these devices is beyond the scope of this paper.
We also note that \citet{schmid2018sphere} and \citet{hunziker2020refplanets} are able to correct the beam shifts of SPHERE-ZIMPOL by measuring them from on-sky data. 
This correction significantly reduces the speckle noise at angular separations ${>}0.6\arcsec$ from the star, but residuals remain at separations ${<}0.6\arcsec$. 
These residuals are particularly strong for broadband data because the beam shifts are wavelength dependent and thus cannot be corrected with a simple shift for a broad wavelength range. 
Therefore, mitigating the beam shifts already during the optical design is the preferred approach.

The size of the spatial GH and IF shifts is independent of the f-number $F$ of the beam of light incident on a mirror.
However, as explained in Sect.~\ref{sec:sgh}, the size of these shifts relative to the size of the PSF is inversely proportional to the f-number.
Therefore, to limit the effect of the beam shifts and the polarization structure they create, the absolute f-numbers of the beams falling onto the mirrors in the optical system should be large; the beams should converge or diverge slowly.
In the limit of a perfectly collimated beam ($F = \infty$), the spatial GH and IF shifts are even negligibly small compared to the size of the PSF.
Because any beam of finite extent corresponds to an angular spectrum of plane waves, the spatial GH and IF shifts still occur for a perfectly collimated beam, but the PSF is located at an infinite distance and is infinitely large.
We note that magnifications in the optical system after the reflection off the mirror do not affect the size of the beam shifts relative to the PSF, because magnifications change the size of the shifts and the PSF by an equal amount.

The spatial GH and IF shifts are created by respectively the phase gradient and the retardance of the mirror at the central angle of incidence of the beam; the amplitudes of the reflection coefficients have only a marginal effect and are therefore not important.
Hence, to minimize the spatial GH and IF shifts, the phase gradient should be kept small and the retardance should have a value close to $180\degr$ (see Eqs.~(\ref{eq:sgh}) and (\ref{eq:sifx})).
Fortunately, the values of the phase gradient and the retardance are closely related: A retardance close to $180\degr$ automatically implies small phase gradients in both the $p$- and $s$-directions.
Figure~\ref{fig:fresnel} (right) shows that this situation occurs at small angles of incidence.
Therefore, to minimize the spatial GH and IF shifts, the central angle of incidence of the beams should be kept small.

Keeping the f-numbers large and the central angles of incidence small may not always be possible because optical systems need to fit in a limited volume.
Therefore, also the design of the coatings of the mirrors should be considered to minimize the spatial GH and IF shifts.
In general, mirror coatings are optimized for large reflectivity to maximize the throughput of the optical system. 
However, highly reflective coatings almost always have retardances significantly different from $180\degr$ and therefore such coatings produce large spatial GH and IF shifts.
But for high-contrast imaging, a high system throughput is of little use when one cannot attain the contrast to image exoplanets. 
Therefore, a paradigm shift in the design of the mirror coatings for high-contrast imagers is necessary: Rather than maximizing the reflectivity, the retardance should be optimized to have values close to $180\degr$ for the central angle of incidence of the mirror and the wavelength range of interest.
For linear polarimeters such a design philosophy has the added advantage that it also prevents large losses of signal due to strong polarimetric crosstalk, such as those found for the image derotators of SPHERE and SCExAO-CHARIS \citep{de2020polarimetric, van2020calibration, van2020polarimetric, thart2021full}.
The larger instrumental polarization resulting from the suboptimal reflectivity is not an issue because it can be easily removed by adding a half-wave plate to the optical path or subtracting it in the data reduction. 


\subsection{Table summarizing properties of beam shifts}
\label{sec:overview_beam_shifts}

In Table~\ref{tab:beam_shifts} we present an overview of the properties of the four beam shifts discussed in this paper.
For each shift, the table shows the type and nature of the effect, the plane or direction of occurrence, the origin of the shift, the parameters that the shift depends on, the typical size, the effect in the focal plane, and whether or not the shift is important for high-contrast imaging. 
Table~\ref{tab:beam_shifts} therefore provides a clear summary of the beam shifts and is a useful reference to compare the effects.

\vspace{1cm}

%
\begin{sidewaystable*}
\caption{Overview of the properties of the four beam shifts discussed in this paper.}
\centering
\resizebox{\vsize}{!}{%
\setlength{\tabcolsep}{7pt}
\renewcommand{\arraystretch}{1.8}
\begin{tabular}{L{3cm} L{5.2cm} L{5.2cm} L{5.2cm} L{5.2cm}}
\hline\hline
Property & Spatial Goos-H\"anchen & Angular Goos-H\"anchen & Spatial Imbert-Federov & Angular Imbert-Federov \vspace{3pt} \\
\hline
Type of shift & Shift of beam of light & Angular deviation of beam of light & Shift of beam of light & Angular deviation of beam of light \\
Nature of effect & Separate shifts for light polarized parallel and perpendicular to plane of incidence & Separate shifts for light polarized parallel and perpendicular to plane of incidence & Separate shifts for diagonally and antidiagonally polarized light as well as for right-handed and left-handed circularly polarized light & Separate shifts for diagonally and antidiagonally polarized light \\
Plane / direction of occurrence & In plane of incidence & In plane of incidence & Perpendicular to plane of incidence & Away from plane of incidence \\
Origin of shift & Variation of Fresnel phases over range of angles of incidence & Variation of Fresnel amplitudes over range of angles of incidence & Conservation of total angular momentum in direction normal to surface of mirror & Conservation of linear momentum in direction perpendicular to plane of incidence \\
Main dependence on Fresnel equations & Phase gradient & Amplitude gradient & Retardance & Diattenuation \\
Dependence on incident Stokes parameters & $Q$ & $Q$ & Primarily $U$ and $V$, weakly $Q$ & Primarily $U$, weakly $Q$ \\
Dependence on f-number of incident beam & No & Yes & No & Yes \\
Typical size & A tenth of a wavelength & 1--100~$\upmu$rad, strongly dependent on f-number of incident beam of light & Less than a tenth of a wavelength & 0.1--10~$\upmu$rad, strongly dependent on f-number of incident beam of light \\
Effect in focal plane for incident unpolarized light & PSF shifted and broadened with underlying polarization structure, in particular splitting of opposite polarizations (ghost PSF) & Basically none & PSF broadened (but not shifted) with underlying polarization structure, in particular splitting of opposite polarizations (ghost PSF) & Basically none \\
Important for high-contrast imaging \vspace{3pt} & Yes & No & Yes & No \\
\hline
\end{tabular}
}
\label{tab:beam_shifts}
\end{sidewaystable*}
%

%
%

\section{Conclusions}
\label{sec:conclusions}

We used polarization ray tracing to numerically compute the beam shifts for reflection off a flat metallic mirror and compared the resulting shifts to the closed-form expressions of the spatial and angular GH and IF shifts from the physics literature.
We find that all four beam shifts are fully reproduced by polarization ray tracing. 
In particular, we find that the phase gradients in the Jones pupil and the ghost PSFs as described by \citet{breckinridge2015polarization} are produced by the spatial GH and IF shifts.
We also studied the origin and characteristics of the four shifts and the dependence of their size and direction on the beam intensity profile, incident polarization state, angle of incidence, mirror material, and wavelength.
An overview of the properties of the four beam shifts is shown in Table~\ref{tab:beam_shifts}.

Whereas the spatial GH and IF shifts depend on the phase of the Fresnel reflection coefficients, the angular GH and IF shifts depend on the amplitude. 
Only the spatial GH and IF shifts are relevant for high-contrast imagers and telescopes because these shifts are visible in the focal plane.
The angular GH and IF shifts on the other hand are not important because they only change the intensity distribution across the reflected beam.
As such, the angular shifts have no significant effect in the focus and only create a small point-symmetric deformation of the PSF.
We thus conclude that only phase aberrations are important; amplitude aberrations have an almost negligible effect.

The spatial GH and IF shifts create a polarization structure in the PSF that reduces the performance of coronagraphs. 
In fact, we find that the polarization structure for the case of unpolarized light incident on a flat metallic mirror is adequately described by the diattenuation (i.e.,~the instrumental polarization) and the spatial GH and IF shifts.
The polarization structure created by the spatial GH and IF shifts can also significantly reduce the speckle suppression of polarimetric measurements, thereby limiting the maximum attainable gain in contrast from polarimetry.
To mitigate the spatial GH and IF shifts in optical systems, the beams of light reflecting off the mirrors should have large f-numbers and small central angles of incidence. 
Most importantly, mirror coatings should not be optimized for maximum reflectivity, but should instead be designed to have a retardance close to $180\degr$.

Our study provides a fundamental understanding of the polarization aberrations resulting from reflection off flat metallic mirrors in terms of beam shifts.
In addition, we have created the analytical and numerical tools to describe these shifts.
The next step is to study the combined effect and wavelength dependence of the beam shifts of complete optical paths of (polarimetric) high-contrast imaging instruments and telescopes with multiple inclined and rotating components, including half-wave plates.
In particular, we plan to use our tools to create a detailed model of the beam shifts affecting the polarimetric mode of SPHERE-ZIMPOL and enable accurate corrections of on-sky observations.
The insights from our work can be applied to understand and improve the performance of many future space- and ground-based high-contrast imagers and polarimeters, such as the Roman Space Telescope, the Habitable Worlds Observatory, GMagAO-X at the Giant Magellan Telescope, PSI at the Thirty Meter Telescope, and PCS (or EPICS) at the Extremely Large Telescope.

%
%

\begin{acknowledgements}

We thank Prof. Dr. Hans Martin Schmid (ETH Zurich) for providing valuable comments on the manuscript. RGvH thanks ESO for the studentship at ESO Santiago during which part of this project was performed. The research of FS and SPB leading to these results has received funding from the European Research Council under ERC Starting Grant agreement 678194 (FALCONER). This research has made use of NASA’s Astrophysics Data System Bibliographic Services; Scipy, a free and open-source Python library used for scientific computing and technical computing \citep{virtanen2020scipy}; Astropy, a community-developed core Python package for Astronomy \citep{robitaille2013astropy, price2018astropy}; and HCIPy, an open-source object-oriented framework written in Python for performing end-to-end simulations of high-contrast imaging instruments \citep{por2018high}.

\end{acknowledgements}

%
%

\bibliographystyle{aa}   
\bibliography{biblio}   

\begin{thebibliography}{57}
\expandafter\ifx\csname natexlab\endcsname\relax\def\natexlab#1{#1}\fi

\bibitem[{Aiello {et~al.}(2009)Aiello, Merano, \& Woerdman}]{aiello2009duality}
Aiello, A., Merano, M., \& Woerdman, J. 2009, Physical Review A, 80, 061801

\bibitem[{Aiello \& Woerdman(2007)}]{aiello2007reflection}
Aiello, A. \& Woerdman, H. 2007, arXiv preprint arXiv:0710.1643

\bibitem[{Aiello \& Woerdman(2008)}]{aiello2008role}
Aiello, A. \& Woerdman, J. 2008, Optics letters, 33, 1437

\bibitem[{Anche {et~al.}(2018)Anche, Anupama, Sriram, Sankarasubramanian, \&
  Skidmore}]{anche2018estimation}
Anche, R.~M., Anupama, G., Sriram, S., Sankarasubramanian, K., \& Skidmore, W.
  2018, in Adaptive Optics Systems VI, Vol. 10703, International Society for
  Optics and Photonics, 107034K

\bibitem[{Anche {et~al.}(2023)Anche, Ashcraft, Haffert, Millar-Blanchaer,
  Douglas, Snik, Williams, van Holstein, Doelman, Van~Gorkom, \&
  Skidmore}]{anche2023polarization}
Anche, R.~M., Ashcraft, J.~N., Haffert, S.~Y., {et~al.} 2023, Astronomy \&
  Astrophysics, 672, A121

\bibitem[{Bliokh \& Aiello(2013)}]{bliokh2013goos}
Bliokh, K.~Y. \& Aiello, A. 2013, Journal of Optics, 15, 014001

\bibitem[{Bliokh \& Bliokh(2006)}]{bliokh2006conservation}
Bliokh, K.~Y. \& Bliokh, Y.~P. 2006, Physical review letters, 96, 073903

\bibitem[{Bliokh \& Bliokh(2007)}]{bliokh2007polarization}
Bliokh, K.~Y. \& Bliokh, Y.~P. 2007, Physical Review E, 75, 066609

\bibitem[{Bliokh \& Nori(2015)}]{bliokh2015transverse}
Bliokh, K.~Y. \& Nori, F. 2015, Physics Reports, 592, 1

\bibitem[{Bliokh {et~al.}(2015)Bliokh, Rodr{\'\i}guez-Fortu{\~n}o, Nori, \&
  Zayats}]{bliokh2015spin}
Bliokh, K.~Y., Rodr{\'\i}guez-Fortu{\~n}o, F.~J., Nori, F., \& Zayats, A.~V.
  2015, Nature Photonics, 9, 796

\bibitem[{Born \& Wolf(2013)}]{born2013principles}
Born, M. \& Wolf, E. 2013, Principles of optics: electromagnetic theory of
  propagation, interference and diffraction of light (Elsevier)

\bibitem[{Breckinridge {et~al.}(2018)Breckinridge, Kupinski, Davis, Daugherty,
  \& Chipman}]{breckinridge2018terrestrial}
Breckinridge, J., Kupinski, M., Davis, J., Daugherty, B., \& Chipman, R. 2018,
  in Space Telescopes and Instrumentation 2018: Optical, Infrared, and
  Millimeter Wave, Vol. 10698, International Society for Optics and Photonics,
  106981D

\bibitem[{Breckinridge {et~al.}(2015)Breckinridge, Lam, \&
  Chipman}]{breckinridge2015polarization}
Breckinridge, J.~B., Lam, W. S.~T., \& Chipman, R.~A. 2015, Publications of the
  Astronomical Society of the Pacific, 127, 445

\bibitem[{Canovas {et~al.}(2011)Canovas, Rodenhuis, Jeffers, Min, \&
  Keller}]{canovas2011data}
Canovas, H., Rodenhuis, M., Jeffers, S., Min, M., \& Keller, C. 2011, Astronomy
  \& Astrophysics, 531, A102

\bibitem[{Chipman(1989)}]{chipman1989polarization}
Chipman, R.~A. 1989, Optical engineering, 28, 280290

\bibitem[{Clark \& Breckinridge(2011)}]{clark2011polarization}
Clark, N. \& Breckinridge, J.~B. 2011, in Uv/Optical/Ir Space Telescopes and
  Instruments: Innovative Technologies and Concepts V, Vol. 8146, International
  Society for Optics and Photonics, 81460O

\bibitem[{Dai {et~al.}(2019)Dai, He, Wang, \& Booth}]{dai2019adaptive}
Dai, Y., He, C., Wang, J., \& Booth, M. 2019, in Adaptive Optics and Wavefront
  Control for Biological Systems V, Vol. 10886, International Society for
  Optics and Photonics, 1088609

\bibitem[{Davis {et~al.}(2018)Davis, Kupinski, Chipman, \&
  Breckinridge}]{davis2018habex}
Davis, J., Kupinski, M.~K., Chipman, R.~A., \& Breckinridge, J.~B. 2018, in
  Space Telescopes and Instrumentation 2018: Optical, Infrared, and Millimeter
  Wave, Vol. 10698, International Society for Optics and Photonics, 106983H

\bibitem[{de~Boer {et~al.}(2020)de~Boer, Langlois, Van~Holstein, Girard,
  Mouillet, Vigan, Dohlen, Snik, Keller, Ginski, {et~al.}}]{de2020polarimetric}
de~Boer, J., Langlois, M., Van~Holstein, R.~G., {et~al.} 2020, Astronomy \&
  Astrophysics, 633, A63

\bibitem[{Espinosa-Luna {et~al.}(2008)Espinosa-Luna, Rodr{\'\i}guez-Carrera,
  Bernabeu, \& Hinojosa-Ru{\'\i}z}]{espinosa2008transformation}
Espinosa-Luna, R., Rodr{\'\i}guez-Carrera, D., Bernabeu, E., \&
  Hinojosa-Ru{\'\i}z, S. 2008, Optik, 119, 757

\bibitem[{Federov(1955)}]{federov1955k}
Federov, F. 1955, in Doklady Akad. Nauk USSR, Vol. 105, 465

\bibitem[{Goos \& H{\"a}nchen(1947)}]{goos1947neuer}
Goos, F. \& H{\"a}nchen, H. 1947, Annalen der Physik, 436, 333

\bibitem[{G{\"o}tte \& Dennis(2012)}]{gotte2012generalized}
G{\"o}tte, J.~B. \& Dennis, M.~R. 2012, New Journal of Physics, 14, 073016

\bibitem[{Hamaker \& Bregman(1996)}]{hamaker1996understanding}
Hamaker, J. \& Bregman, J. 1996, Astronomy and Astrophysics Supplement Series,
  117, 161

\bibitem[{Heiles(2000)}]{heiles20009286}
Heiles, C. 2000, The Astronomical Journal, 119, 923

\bibitem[{Hermosa {et~al.}(2011)Hermosa, Nugrowati, Aiello, \&
  Woerdman}]{hermosa2011spin}
Hermosa, N., Nugrowati, A., Aiello, A., \& Woerdman, J. 2011, Optics letters,
  36, 3200

\bibitem[{Hunziker {et~al.}(2020)Hunziker, Schmid, Mouillet, Milli, Zurlo,
  Delorme, Abe, Avenhaus, Baruffolo, Bazzon, {et~al.}}]{hunziker2020refplanets}
Hunziker, S., Schmid, H.~M., Mouillet, D., {et~al.} 2020, Astronomy \&
  Astrophysics, 634, A69

\bibitem[{Imbert(1972)}]{imbert1972calculation}
Imbert, C. 1972, Physical Review D, 5, 787

\bibitem[{Krist {et~al.}(2017)Krist, Riggs, McGuire, Tang, Amiri, Gutt,
  Marchen, Marx, Nemati, Saini, {et~al.}}]{krist2017wfirst}
Krist, J., Riggs, A., McGuire, J., {et~al.} 2017, in Techniques and
  Instrumentation for Detection of Exoplanets VIII, Vol. 10400, International
  Society for Optics and Photonics, 1040004

\bibitem[{Mawet {et~al.}(2014)Mawet, Milli, Wahhaj, Pelat, Absil, Delacroix,
  Boccaletti, Kasper, Kenworthy, Marois, {et~al.}}]{mawet2014fundamental}
Mawet, D., Milli, J., Wahhaj, Z., {et~al.} 2014, The Astrophysical Journal,
  792, 97

\bibitem[{McGuire \& Chipman(1990)}]{mcguire1990diffraction}
McGuire, J.~P. \& Chipman, R.~A. 1990, JOSA A, 7, 1614

\bibitem[{McGuire \& Chipman(1994{\natexlab{a}})}]{mcguire1994polarization1}
McGuire, J.~P. \& Chipman, R.~A. 1994{\natexlab{a}}, Applied optics, 33, 5080

\bibitem[{McGuire \& Chipman(1994{\natexlab{b}})}]{mcguire1994polarization2}
McGuire, J.~P. \& Chipman, R.~A. 1994{\natexlab{b}}, Applied optics, 33, 5101

\bibitem[{Mendillo {et~al.}(2019)Mendillo, Howe, Hewawasam, Martel, Cook, \&
  Chakrabarti}]{mendillo2019polarization}
Mendillo, C.~B., Howe, G.~A., Hewawasam, K., {et~al.} 2019, Journal of
  Astronomical Telescopes, Instruments, and Systems, 5, 025003

\bibitem[{Merano {et~al.}(2007)Merano, Aiello, Van~Exter, Eliel, Woerdman,
  {et~al.}}]{merano2007observation}
Merano, M., Aiello, A., Van~Exter, M., {et~al.} 2007, Optics express, 15, 15928

\bibitem[{Millar-Blanchaer {et~al.}(2022)Millar-Blanchaer, Anche, Nguyen, \&
  Maire}]{millar2022polarization}
Millar-Blanchaer, M.~A., Anche, R.~M., Nguyen, M.~M., \& Maire, J. 2022, in
  Ground-based and Airborne Instrumentation for Astronomy IX, Vol. 12184, SPIE,
  1278--1288

\bibitem[{Millar-Blanchaer {et~al.}(2016)Millar-Blanchaer, Perrin, Hung,
  Fitzgerald, Wang, Chilcote, Graham, Bruzzone, \& Kalas}]{millar2016gpi}
Millar-Blanchaer, M.~A., Perrin, M.~D., Hung, L.-W., {et~al.} 2016, in
  Ground-based and Airborne Instrumentation for Astronomy VI, Vol. 9908,
  International Society for Optics and Photonics, 990836

\bibitem[{Por {et~al.}(2018)Por, Haffert, Radhakrishnan, Doelman, van Kooten,
  \& Bos}]{por2018high}
Por, E.~H., Haffert, S.~Y., Radhakrishnan, V.~M., {et~al.} 2018, in Adaptive
  Optics Systems VI, Vol. 10703, International Society for Optics and
  Photonics, 1070342

\bibitem[{Price-Whelan {et~al.}(2018)Price-Whelan, Sip{\H{o}}cz, G{\"u}nther,
  Lim, Crawford, Conseil, Shupe, Craig, Dencheva, Ginsburg,
  {et~al.}}]{price2018astropy}
Price-Whelan, A.~M., Sip{\H{o}}cz, B., G{\"u}nther, H., {et~al.} 2018, The
  Astronomical Journal, 156, 123

\bibitem[{Raki{\'c} {et~al.}(1998)Raki{\'c}, Djuri{\v{s}}i{\'c}, Elazar, \&
  Majewski}]{rakic1998optical}
Raki{\'c}, A.~D., Djuri{\v{s}}i{\'c}, A.~B., Elazar, J.~M., \& Majewski, M.~L.
  1998, Applied optics, 37, 5271

\bibitem[{Robitaille {et~al.}(2013)Robitaille, Tollerud, Greenfield,
  Droettboom, Bray, Aldcroft, Davis, Ginsburg, Price-Whelan, Kerzendorf,
  {et~al.}}]{robitaille2013astropy}
Robitaille, T.~P., Tollerud, E.~J., Greenfield, P., {et~al.} 2013, Astronomy \&
  Astrophysics, 558, A33

\bibitem[{Sabatke {et~al.}(2018)Sabatke, Knight, \&
  Bolcar}]{sabatke2018polarization}
Sabatke, D., Knight, J.~S., \& Bolcar, M.~R. 2018, in Optical Modeling and
  Performance Predictions X, Vol. 10743, International Society for Optics and
  Photonics, 1074307

\bibitem[{Safonov {et~al.}(2019)Safonov, Lysenko, Goliguzova, \&
  Cheryasov}]{safonov2019differential}
Safonov, B., Lysenko, P., Goliguzova, M., \& Cheryasov, D. 2019, Monthly
  Notices of the Royal Astronomical Society, 484, 5129

\bibitem[{Sanchez~Almeida \& Martinez~Pillet(1992)}]{sanchez1992instrumental}
Sanchez~Almeida, J. \& Martinez~Pillet, V. 1992, Astronomy and Astrophysics,
  260, 543

\bibitem[{Schmid {et~al.}(2018)Schmid, Bazzon, Roelfsema, Mouillet, Milli,
  Menard, Gisler, Hunziker, Pragt, Dominik, {et~al.}}]{schmid2018sphere}
Schmid, H.~M., Bazzon, A., Roelfsema, R., {et~al.} 2018, Astronomy \&
  Astrophysics, 619, A9

\bibitem[{Sit {et~al.}(2017)Sit, Giner, Karimi, \& Lundeen}]{sit2017general}
Sit, A., Giner, L., Karimi, E., \& Lundeen, J.~S. 2017, Journal of Optics, 19,
  094003

\bibitem[{'t~Hart {et~al.}(2021)'t~Hart, van Holstein, Bos, Ruigrok, Snik,
  Lozi, Guyon, Kudo, Zhang, Jovanovic, {et~al.}}]{thart2021full}
't~Hart, J. G.~J., van Holstein, R.~G., Bos, S.~P., {et~al.} 2021, in
  Polarization Science and Remote Sensing X, Vol. 11833, SPIE, 148--174

\bibitem[{Totzeck {et~al.}(2005)Totzeck, Graupner, Heil, Gohnermeier, Dittmann,
  Krahmer, Kamenov, Ruoff, \& Flagello}]{totzeck2005describe}
Totzeck, M., Graupner, P., Heil, T., {et~al.} 2005, in Optical Microlithography
  XVIII, Vol. 5754, International Society for Optics and Photonics, 23--37

\bibitem[{van Holstein {et~al.}(2020{\natexlab{a}})van Holstein, Bos, Ruigrok,
  Lozi, Guyon, Norris, Snik, Chilcote, Currie, Groff,
  {et~al.}}]{van2020calibration}
van Holstein, R.~G., Bos, S.~P., Ruigrok, J., {et~al.} 2020{\natexlab{a}}, in
  Ground-based and Airborne Instrumentation for Astronomy VIII, Vol. 11447,
  International Society for Optics and Photonics, 114475B

\bibitem[{van Holstein {et~al.}(2020{\natexlab{b}})van Holstein, Girard,
  De~Boer, Snik, Milli, Stam, Ginski, Mouillet, Wahhaj, Schmid,
  {et~al.}}]{van2020polarimetric}
van Holstein, R.~G., Girard, J.~H., De~Boer, J., {et~al.} 2020{\natexlab{b}},
  Astronomy \& Astrophysics, 633, A64

\bibitem[{Virtanen {et~al.}(2020)Virtanen, Gommers, Oliphant, Haberland, Reddy,
  Cournapeau, Burovski, Peterson, Weckesser, Bright,
  {et~al.}}]{virtanen2020scipy}
Virtanen, P., Gommers, R., Oliphant, T.~E., {et~al.} 2020, Nature methods, 17,
  261

\bibitem[{Waluschka(1989)}]{waluschka1989polarization}
Waluschka, E. 1989, Optical Engineering, 28, 280286

\bibitem[{Wiktorowicz {et~al.}(2014)Wiktorowicz, Millar-Blanchaer, Perrin,
  Graham, Fitzgerald, Maire, Ingraham, Savransky, Macintosh, Thomas,
  {et~al.}}]{wiktorowicz2014gemini}
Wiktorowicz, S.~J., Millar-Blanchaer, M., Perrin, M.~D., {et~al.} 2014, in
  Ground-based and Airborne Instrumentation for Astronomy V, Vol. 9147,
  International Society for Optics and Photonics, 914783

\bibitem[{Will \& Fienup(2019)}]{will2019effects}
Will, S.~D. \& Fienup, J.~R. 2019, in Techniques and Instrumentation for
  Detection of Exoplanets IX, Vol. 11117, International Society for Optics and
  Photonics, 1111710

\bibitem[{Witzel {et~al.}(2011)Witzel, Eckart, Buchholz, Zamaninasab, Lenzen,
  Sch{\"o}del, Araujo, Sabha, Bremer, Karas, {et~al.}}]{witzel2011instrumental}
Witzel, G., Eckart, A., Buchholz, R., {et~al.} 2011, Astronomy \& Astrophysics,
  525, A130

\bibitem[{Yun {et~al.}(2011{\natexlab{a}})Yun, Crabtree, \&
  Chipman}]{yun2011three1}
Yun, G., Crabtree, K., \& Chipman, R.~A. 2011{\natexlab{a}}, Applied optics,
  50, 2855

\bibitem[{Yun {et~al.}(2011{\natexlab{b}})Yun, McClain, \&
  Chipman}]{yun2011three2}
Yun, G., McClain, S.~C., \& Chipman, R.~A. 2011{\natexlab{b}}, Applied optics,
  50, 2866

\end{thebibliography}

%
%

\begin{appendix}

\onecolumn

\section{Amplitude-response matrix}
\label{app:arm}

\begin{figure*}[!htpb]
\centering
\includegraphics[width=\hsize, trim={7 7 7 5}, clip]{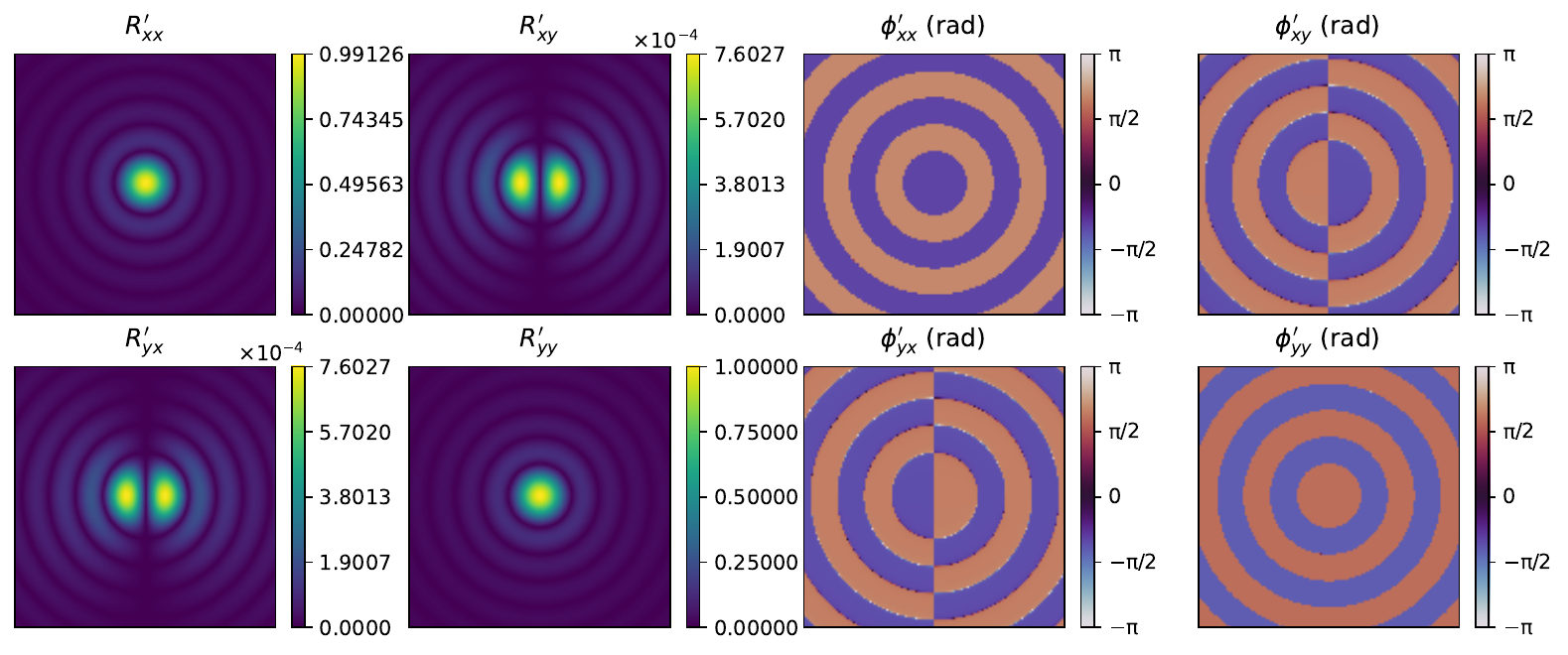} 
\caption{Amplitude-response matrix (ARM) expressed in the $xyz$-basis at a wavelength of 820~nm for a converging beam of light with an f-number of 61.3 that reflects off gold at an angle of incidence of $45\degr$. The panels in the first and second (third and fourth) columns show the amplitude (phase) of the central $500~\upmu$m $\times$ $500~\upmu$m of the ARM elements. The positive $x$- and $y$-directions are upward and to the left, respectively. The values of the color maps are different among the panels in the first and second columns.}
\label{fig:arm} 
\end{figure*}

Figure~\ref{fig:arm} shows the amplitude-response matrix for reflection with an angle of incidence of $45\degr$.

%

\end{appendix}

\end{document}